\documentclass[12pt]{article}
\usepackage{a4,amsmath,amssymb,cite,graphicx}
\def\Bbb{\mathbb}
\def\BZ{\Bbb Z} \def\BR{\Bbb R}
\def\BC{\Bbb C} \def\BP{\Bbb P}

\catcode`@=11 \@addtoreset{equation}{section} \catcode`@=12

\newcommand{\ch}{\textrm{ch}}

\begin{document}
\begin{titlepage}
\begin{flushleft}
{\tt hep-th/0504164}
\hfill April 2005(v1) \\
{\tt IITM/2005/PH/TH/2}\hfill June 2005(v2) 
\end{flushleft}
\begin{center}
{\Large \bf  A quantum McKay correspondence for fractional $2p$-branes
on LG orbifolds
} \\[1cm]
Bobby Ezhuthachan\\
{\em The Institute of Mathematical Sciences, \\ Chennai 600 113, India\\
{\tt Email: bobby@imsc.res.in}\\[5pt]}
\mbox{} \\
Suresh Govindarajan\\
{\em Department of Physics, Indian Institute of Technology, Madras,\\
Chennai 600 036, India\\
{\tt Email: suresh@physics.iitm.ac.in}\\[5pt]}
and \\[5pt]
T. Jayaraman\footnote{On leave of absence 
from The Institute of Mathematical Sciences,
Chennai}
\\ {\em Dept. of Mathematics, \\ Tata Institute of
Fundamental Research,\\ Mumbai 400 005 India}\\
{\tt Email: jayaram@theory.tifr.res.in}
\end{center}
\vfill
\begin{abstract}
We study fractional $2p$-branes and their intersection numbers in
non-compact orbifolds as well the continuation of these objects
in K\"ahler moduli space to coherent sheaves in the corresponding
smooth non-compact Calabi-Yau manifolds. We show that the
restriction of these objects to compact Calabi-Yau hypersurfaces
gives the new fractional branes in LG orbifolds constructed by
Ashok {\em et. al.} in {\tt hep-th/0401135}. We thus demonstrate the
equivalence of the B-type branes corresponding to linear boundary
conditions in LG orbifolds, originally constructed in {\tt
hep-th/9907131}, to a subset of those constructed in LG orbifolds
using boundary fermions and matrix factorization of the
world-sheet superpotential. The relationship between the coherent
sheaves corresponding to the fractional two-branes leads to a
generalization of the McKay correspondence that we call the
quantum McKay correspondence due to a close parallel with the
construction of branes on non-supersymmetric orbifolds.  We also
provide evidence that the boundary states associated to these
branes in a conformal field theory description corresponds to a
sub-class of the boundary states associated to the permutation
branes in the Gepner model associated with the LG orbifold.
\end{abstract}
\vfill
\end{titlepage}

\section{Introduction}

During the past five years, our understanding of the spectrum of
D-branes (both A and B-type)
that appear in type II compactifications on Calabi-Yau(CY)
manifolds has significantly improved. Some of the progress has been
achieved by relating geometric constructions to non-geometric ones such
as LG orbifolds and Gepner models. The first step in this context
appeared in the \cite{quintic} (see also \cite{gjs} and \cite{mikeictp}
for a review). The method
proposed for B-type D-branes
was to analytically continue periods (identified with the
central charge of D-branes) from non-geometric regions to geometric
regions and then use this information to
add geometric insight into  the story. This method is rather tedious but
it lead the way to a simpler picture for the large-volume analog of
the Recknagel-Schomerus(RS)
boundary states in the Gepner model\cite{rs}. From this
emerged a connection with the McKay
correspondence\cite{dd,mckay,tomas,mayr,Lerche:2001vj}. The RS boundary states turned
out to be restriction of vector bundles or more
generally coherent sheaves on the ambient (weighted) projective space
to the CY hypersurface.  However there are still several important 
aspects that are unclear and need to be clarified further. Among these is 
the question of 
an explicit description of B-type D-branes on Calabi-Yau manifolds
that are not of this type. 

A significant recent development in this direction 
has been the study of B-type D-branes at the Landau-Ginzburg(LG) point 
in the K\"ahler moduli space of a Calabi-Yau manifold. As one might, with
hindsight, have expected from the simplicity of the  LG
theory, it is indeed possible to provide a fairly explicit description of 
B-type branes using boundary fermions and the technique of matrix factorization
of polynomials\cite{kapustinli}. This construction follows closely the conjecture 
of Kontsevich regarding the categorical description of B-type branes in the LG
theory. In an interesting 
development it was shown in \cite{Ashok} that a new class of fractional
branes can in fact be defined in the LG theory. The D-brane charges of 
these objects in terms of the charge basis at the large-volume point 
in K\"ahler moduli space have been computed. 

Interestingly these fractional branes (in the LG description)
include an object that corresponds at large volume to a single 
zero brane on the  CY manifold. This is of particular interest since the 
Recknagel-Schomerus construction of boundary states in the Gepner
model for CY manifolds appears to generically miss
the D0-brane on the CY manifold. This D0-brane together with others that
are related to it by the quantum symmetry at the LG point are of course
only some examples of a large class of new branes that can be constructed 
using the technique of matrix factorization of the world-sheet superpotential
of the LG Lagrangian. 
    
This new approach to B-type branes in the 
LG theory is due to several authors
\cite{kapustinli,Ashok,horiwalcher,lerchewalcher,walcher}.
For completeness we may mention 
another major development in this new approach
has been the computation of 
the world-volume superpotential of such branes\cite{Ashok2}. We also note 
that, from a purely mathematical point of view, there have been further 
developments in the categorical description of these B-type branes 
at the LG point and the derived equivalence of this category
to the derived category of coherent sheaves on the same CY at large 
volume appears to have been established.
This has been done in a series of papers by Orlov \cite{orlov}.

In this paper we will investigate the new fractional branes, referred to
above, due to ref \cite{Ashok}, by different methods without using 
the technique of matrix factorization and the introduction of 
boundary fermions. 
Our aim in particular, is to  
understand these states from a more transparently geometric point of 
view. 

The key observation is that the new fractional branes essentially 
arise from considering Neumann type boundary conditions on the fields
of the LG theory. In an earlier paper\cite{gjs} it was observed that there
were constraints on how Neumann boundary conditions could be imposed 
on the fields in the LG model in order to describe
D-branes\footnote{This work is a consequence of our attempt to relate
these two apparently distinct constructions.}. There 
was a {\em consistency condition} involving the world-sheet superpotential. 
In the quintic we observed, for instance, that Neumann boundary
conditions  could be imposed
only on linear combinations  of LG fields and not on the individual
fields themselves. This is very similar to the particular matrix factorization
that ref. \cite{Ashok}) use in
order to construct their new fractional branes.

Thus instead of the old fractional branes, which were fractional zero-branes,
located at the singular point, we now consider fractional two-branes which
are complex lines that pass through the fixed point. One may easily construct
such objects in the orbifold theory without the world-sheet superpotential.
The D-brane charges
of these objects in the large volume basis can be computed by first determining
the intersection of these fractional two-branes with the fractional zero-branes
and with themselves. This  can be easily done at the orbifold point.
Subsequently, using these intersection numbers we can determine the 
large-volume charges of these fractional two-branes.
It turns out that these fractional two-branes have a `fractional' first 
Chern class. This fractional first Chern class ensures that 
when these objects are restricted to the CY hypersurface, one of them 
precisely becomes a zero-brane on the CY hypersurface. One may note
that if we began with objects that have integer first Chern class on the 
ambient projective
space then we would always obtain $d$ zero-branes on the CY hypersurface,
where $d$ is the degree of the polynomial equation describing the CY.  

Since the appearance of a `fractional' first Chern class is somewhat 
surprising, we also show that these fractional two-branes have integer
charges if we consider them as objects living in the ambient non-compact
CY, which is a line bundle over a projective base,
rather than on the projective space itself. In the Gauged
Linear Sigma Model (GLSM)
description, the true ambient space provided by the fields of the theory
is in fact a non-compact CY. The CY itself comes from first restricting 
to the (weighted) projective space that is the base of the non-compact CY,
and then restricting it to the appropriate hypersurface. 

We also show that all the boundary states in the Ramond sector
corresponding to the fractional two-branes can be described using 
the boundary conditions on the {\em bulk} world-sheet fermions. It was 
observed in \cite{mayr} (see also \cite{bundlesglsm})
that this in fact could be done for the fractional
zero-branes themselves and in this paper we extend this observation to 
the new fractional branes.   

For definiteness, we illustrate our method mainly in the case of the 
non-compact orbifold $\BC^3/\BZ_3$, the corresponding CY hypersurface being 
the elliptic curve given by a degree three equation in $\BP^2$. The extension 
to the case of the quintic is straightforward. We note also that 
these fractional two-branes in non-compact orbifolds were earlier 
considered by Romelsberger\cite{romel1} 
(for a discussion of the related boundary
state construction see also \cite{BCR}) and 
the appearance of a `fractional' first Chern class was noted indirectly.
This was explained there by the interplay of the relative homology
of the ambient non-compact CY and the compact homology of the base
projective space. Our description of the fractional two-branes in the 
ambient non-compact CY provides a clear toric description of the same 
phenomenon. 

In this paper we also note a strong parallel between the relation of the 
fractional two-branes (and more generally fractional $2p$-branes) in the orbifold
theory to the corresponding coherent sheaves at large volume and the notion
of the quantum McKay correspondence due to Martinec and Moore \cite{MM}. 
While there is nothing quantum about the relation in our setting where 
space-time supersymmetry is not broken, nevertheless the fractional 
two-branes in $\BC^n/\BZ_N$ appear closely related to the fractional
zero-branes of $\BC^{n-1}/\BZ_N$, the latter being the geometry associated
with the non-supersymmetric B-type branes in the work of Martinec and Moore.
In the full set of coherent sheaves that correspond to the quantum $\BZ_N$
orbit of the fractional two-branes at large volume we are able to find the 
analogues of the so-called `Coulomb branes' that they describe.\footnote{The 
authors of this paper had of course the choice of referring to 
the analogue of the quantum McKay correspondence in the case at hand, of 
supersymmetric fractional-$2p$ branes, by a different name, possibly as an 
`extended McKay correspondence'. However 
in order not to further increase jargon for what is a closely related 
geometric phenomenon we retain the nomenclature developed in
\cite{MM}.}

Finally we also identify the CFT description of states
of the LG model that correspond to fractional two-brane and four-brane 
states in the ambient non-compact orbifold that are restricted to the CY 
hypersurface. This turns out to be a sub-class of the B-type 
permutation branes of 
the Gepner models that have been studied earlier\cite{permutation}.
  
The organization of the paper is as follows: 
In section 2, we discuss the background to the paper and explain the
setting of the problem. In particular, we focus on linear boundary
conditions in LG orbifolds. Using the GLSM,
section 3 gives a heuristic derivation of
the large volume analogs of D-branes for a specific set of boundary 
conditions in the LG orbifold for the Fermat quintic. We provide
evidence that these are indeed the new fractional branes obtained in
the first reference\cite{Ashok}.
In section 4, we focus on fractional
$2p$-branes in orbifolds of $\BC^n/\BZ_N$, in order to better understand
the results of section 3. 
We work out a
{\em master} formula, Eq. (\ref{master}),
for the intersection forms at the orbifold. These
intersection forms are independently computed in the large volume
as well. The case of $\BC^3/\BZ_3$ is worked out in some detail. 
Section 5 discusses how, for instance, fractional two-branes in
a supersymmetric orbifold $\BC^n/\BZ_N$ are related to
fractional zero-branes on a related non-supersymmetric orbifold
$\BC^{n-1}/ \BZ_N$. We argue for the existence of
a quantum McKay correspondence which relates sheaves associated with
fractional $2p$-branes on $\BC^n/\BZ_N$ to the sheaves
associated with tautological bundles for 
$(2n\!-\!2p)$-branes on the same orbifold.
Section 6 connects our results in the LG orbifold with boundary
conformal field
theory. We provide evidence that the fractional two-branes and a certain
class of fractional four-branes on restriction to the compact Calabi-Yau
hypersurface are given by a sub-class of the permutation branes
constructed in the Gepner model\cite{permutation}. We conclude in
section 7 with a
summary of our results and some comments on unresolved issues. The
appendices have been used to collect several technical results.

While this work was being readied for publication, a paper
\cite{Brunner:2005fv} appeared and has substantial overlap with the results in
section 6.

\section{Background}

\subsection{Gepner models, LG orbifolds and the GLSM}

Consider a Gepner model obtained by tensoring of $n$ $N=2$ minimal models
of level $k_i$, $i=1,\ldots,n$. The total central charge of the model is
$$
\hat{c} = \sum_{i=1}^n \frac{k_i}{k_i+2} = m
$$

The related LG orbifold is obtained by considering $n$ chiral
superfields(notation as in refs. \cite{wittenphases,glsm,dd})
$\Phi_i$ with a $\BZ_K$ action ($K\equiv\textrm{lcm}(k_1,\ldots,k_n)$):
$$
\BZ_K:\quad \Phi_i \rightarrow \omega^{Q_i}\ \Phi_i\quad,
$$
where $\omega=\exp(i2\pi/K)$ is a $K$-th root of unity and 
$Q_i\equiv K/(k_i+2)$. The model has a quasi-homogeneous superpotential
given by
\begin{equation}
G(\Phi)=\sum_{i=1}^{n} \Phi^{k_i+2}\ .
\label{lgpot}
\end{equation}
We will be focusing mainly on the case when $n=m+2$ for which $K=\sum_i Q_i$. 

The LG orbifold can be obtained as a `phase' of a gauged linear sigma
model (GLSM) by coupling the 
chiral fields to an 
abelian vector multiplet as well as another chiral multiplet $P$.
The charges of the $n$ chiral multiplets under this $U(1)$ are $Q_i$
and the chiral multiplet $P$ has charge $-K$. The D-term constraint 
is
\begin{equation}
\sum_{i=1}^n Q_i |\phi_i|^2 - K |p|^2 = r\ ,
\end{equation}
where $r$ is the Fayet-Iliopoulos(FI) parameter. The superpotential in the
GLSM is taken to be
$$
W(\Phi,P)=P\ G(\Phi)\ ,
$$
where $G(\Phi)$ as defined in Eq. (\ref{lgpot}).

When $r\ll0$, the $p$
field is non-zero in the ground state and the $U(1)$ is broken to $\BZ_K$.
Further, the fluctuations of the $p$ field are massive and can be integrated
out to obtain the LG orbifold. When $r\gg0$, not all the $\phi_i$ can
vanish simultaneously and $p=0$ in the ground state. The $\phi_i$
become homogeneous coordinates on a weighted projective space $\BP^{n-1}[Q_i]$
with weights $Q_i$. The equation of motion of the $p$ field imposes the 
restriction that the fields $\phi_i$ lie on the hypersurface $G=0$ in
$\BP^{n-1}[Q_i]$. In the absence of the superpotential, one obtains
the total space of the line bundle ${\cal O}(-K)$ as the moduli space
with the $\BP^{n-1}[Q_i]$ being given by the condition $p=0$. i.e.,
the zero section of the line bundle.

\subsection{Linear boundary conditions in LG orbifolds}

Now consider the LG orbifold on a worldsheet
with boundary preserving $B$-type supersymmetry. 
Such boundary conditions were studied in \cite{gjs},
where it was shown that linear boundary conditions
are specified by a hermitian matrix $B$ which squares to identity, $B^2=1$
and is block diagonal (it mixes fields with identical  charges).
The boundary conditions then take the form
\begin{eqnarray}
{\big(P_D\big)_i}^j\ \phi_j =0 &,&
{\big(P_D\big)_i}^j\ \tau_j =0 \nonumber \\
{\big(P_N\big)_i}^j\ \xi_j =0 &,&
{\big(P_N\big)_i}^j\ \partial_n\phi_j =0 
\label{genbc}
\end{eqnarray}
where $\xi_i \equiv (\psi_{+i} + \psi_{-i})/\sqrt2$ and $\tau_i \equiv 
(\psi_{+i} - \psi_{-i})/\sqrt2$. The matrices 
$P_N \equiv (1+B)/2$ and $P_D\equiv (1-B)/2$ project onto the
Neumann and Dirichlet directions respectively. 
The matrix $B$ which specifies the boundary conditions needs to satisfy
an additional condition due to the presence of the  
superpotential in the LG model\footnote{It is easy to see 
that this condition along with  ${(P_D)_i}^j\ \phi_j =0$
and the block-diagonal nature of $B$
implies the condition $G=0$ for quasi-homogeneous
superpotentials\cite{gzero}: 
$$
G(\phi)\propto \sum_i Q_i \phi_i \frac{\partial G}{\partial \phi_i}
= \sum_{i,j}  \frac{\partial G}{\partial \phi_i} {(P_N + P_D)_i}^j Q_j \phi_j
= \sum_{i,j}  \frac{\partial G}{\partial \phi_i} {(P_D)_i}^j Q_j \phi_j
= \sum_{i,j}  Q_i \frac{\partial G}{\partial \phi_i} {(P_D)_i}^j  \phi_j
=0\ .
$$
 }
\begin{equation}
\frac{\partial G}{\partial \phi_i}\ {\big(P_N\big)_i}^j  =0\ .
\label{superbc}
\end{equation}

In simple models involving a single chiral field, the only possible condition
is the Dirichlet one\footnote{This assumes the absence of degrees of freedom
other than those that come from the bulk LG theory.}.
This carries over to the case of several chiral
superfields when one imposes boundary conditions separately on each of
the chiral superfields, i.e., the matrix $B$ is taken to be diagonal. 
For LG orbifolds associated with
Gepner models, this implies
that all the boundary states constructed by Recknagel and Schomerus in
\cite{rs}
must necessarily arise from
Dirichlet conditions being imposed on all the chiral superfields.
Further, when the superpotential is degenerate at $\phi_i=0$, the
condition $G=0$ implies that the RS states arise from the
boundary condition $\phi_i=0$ for all $i$. 

\subsection{Relating to D-branes at large volume}

Based on the results for the $\BC^3/\BZ_3$ orbifold\cite{dfr2}, it
was conjectured in \cite{dd} that the fractional
zero-branes on $\BC^n/\Gamma$ (with $\Gamma$ a discrete abelian subgroup of 
$SU(n)$) 
correspond at large volume to (exceptional) coherent sheaves that 
provide a natural basis for
bundles on the exceptional divisor of the (possibly partial) resolution  of
$\BC^n/\Gamma$. A second conjecture in \cite{dd} was that the RS boundary
states (to be precise, the $L_i=0$ RS states) were given by
the restriction of the fractional zero-branes to the Calabi-Yau hypersurface.
Substantial evidence for this was provided in refs. \cite{mckay,tomas,mayr}.  

While the exceptional coherent sheaves obtained from fractional zero-branes
do provide a basis for
sheaves on the exceptional divisor of the resolution, this does not
remain true on restriction to the Calabi-Yau threefold. For the case of
the quintic, one finds that the bundles that
are obtained by restriction span an index-25 sub-lattice of the lattice
of RR charges of the quintic. In particular, the zero-brane and two-brane
charges appear in multiples of 5 of the smallest possible
value\cite{dd,bd}.
This suggests that one must generalise the Recknagel-Schomerus construction
to obtain new boundary states in the Gepner model or equivalently consider
more general boundary conditions in the LG orbifold. As was explained in
the previous subsection, the RS boundary states correspond to 
imposing Dirichlet
boundary conditions on all the fields
in the LG orbifold. It is thus natural  
to consider boundary conditions that impose Neumann boundary conditions
on one or more linear combinations of fields as given in Eq. (\ref{genbc}). 
For instance, for the 
LG orbifold for the superpotential given by the Fermat quintic,
one such boundary condition is (see sec. 3.3.3 of \cite{gjs}) 
\begin{eqnarray}
(\phi_1 + \phi_2)=0&,& \phi_i=0\quad\textrm{for }i=3,4,5\ , \nonumber \\
(\xi_1-\xi_2)&=&0
\label{quinticbc}
\end{eqnarray}
It is easy to see that the above boundary conditions satisfy the constraint
(\ref{superbc}) or equivalently that $G=0$ on the boundary. Such branes
will be referred to as {\em fractional two-branes}. We will see that
restriction of one of the fractional two-branes to the quintic
hypersurface leads to a zero-brane of minimal charge.

\section{Fractional two-branes: the quintic in $\BP^4$}

In this section, we will provide a {\em heuristic} derivation of
the identity, at large volume, of the D-branes associated with the
boundary conditions given in Eq. (\ref{quinticbc}). As with any
heuristic derivation, we will not provide complete justification but
only a motivation for some of the steps involved. Nevertheless, at
the end of the derivation, we will have concrete candidates for
the identity of the new D-branes. These new D-branes will be shown to
have the same intersection numbers as those of the new fractional branes
proposed in \cite{Ashok}. We will also provide further justification
for the heuristic derivation in the  section 4. This, we believe,
provides {\em geometric insight} into the categorical construction of
\cite{Ashok} in addition to identifying two apparently distinct
approaches to D-branes on LG orbifolds. For purposes of illustration, 
we first consider the case of the original fractional zero-branes.

\subsection{Fractional zero-branes from Euler sequences}

As has already been mentioned, the original fractional branes (to be
identified with the Recknagel-Schomerus states) are obtained by
imposing Dirichlet boundary conditions on all fields in the LG orbifold. 
This leaves us with
{\em five} independent fermionic multiplets on the boundary (the bulk
chiral multiplet splits up into a fermionic multiplet and a 
chiral multiplet on the boundary) 
We will use the GLSM to interpolate between the LG orbifold and the
nonlinear sigma model(NLSM).

The simplest way to obtain
the LG orbifold from the GLSM is to consider the limit 
$e^2 r \rightarrow -\infty$. In this limit, the fields in the vector
multiplet behave as Lagrange multipliers. The $D$-field imposes
the $D$-term constraints and the gauginos impose the 
constraint\cite{glsm}:
$$
\sum_i Q_i \phi_i \bar{\psi}_{\pm i}=0\ .
$$
For the case of the LG model with boundary, when one imposes {\em Dirichlet}
boundary conditions on all fields, this equation 
imposes a condition on the fermionic
combination $ \bar{\xi}_i $ 
that is not set to zero 
by the boundary conditions. Thus, the gaugino constraint on the boundary
is now 
\begin{equation}
\sum_i Q_i \phi_i \bar{\xi}_i =0
\label{gaugino}
\end{equation}

It is important to what follows that these fermions $\xi_i$ play the role
of the boundary fermions
that are used to construct coherent sheaves associated
with B-type branes at large volume. The argument for this is based on two
observations.  
First, the six-brane is one of the Recknagel-Schomerus
states and hence arises from having Dirichlet boundary conditions on all
the fields in the case
of the LG orbifold. This implies that in analytically
continuing (in K\"ahler moduli space) the D-branes from
the LG orbifold to the large volume limit,
all Dirichlet boundary conditions become Neumann
boundary conditions (see also the discussion in
\cite{glsm}). Second, in the GLSM construction
that realises B-type branes as coherent sheaves, 
the boundary condition  at large
volume relates the $\xi_i$ to the boundary fermions $\pi_a$
via the boundary 
condition\cite{bundlesglsm}
\begin{equation}
\bar{\xi}_i = i \frac{\partial J^a}{\partial \phi_i} \pi_a
\end{equation}
$J^a(\phi)$ ($a=1,\ldots,r$) are $r$ homogeneous polynomials of degree $d$.
This boundary condition appears for the 
coherent sheaf given by the following exact sequence
$$
0\rightarrow E \rightarrow {\cal O}^{\oplus r} \stackrel{J}{\rightarrow}
{\cal O}(d) \rightarrow 0 
$$
which corresponds to imposing the {\em holomorphic
constraint} $J^a \pi_a=0$ on $r$ fermions, $\pi_a$ ($a=1,\ldots,r$). 
considered as sections of ${\cal O}^{\oplus r}$.
Treating the gaugino constraint in Eq. (\ref{gaugino}) as a degree one
holomorphic constraint, that is setting $J^a \equiv \phi^a$ we obtain the 
Euler sequence on $\BP^4$
$$
0\rightarrow \Omega(1) \rightarrow {\cal O}^{\oplus 5} \rightarrow
{\cal O}(1) \rightarrow 0
$$
with the boundary condition
$$
\bar{\xi}_i = i \pi_i\ .
$$
Thus, one can indeed bypass the introduction of boundary fermions by treating
the $\bar{\xi}_i$ as boundary fermions
and the gaugino constraint as a holomorphic constraint\footnote{This is
also a hint why the matrix factorisation used in \cite{Ashok} must be
equivalent to the boundary conditions considered in \cite{gjs}, at
least for the case of linear factors.}.

Of course, we have five boundary states associated with the LG orbifold.
It turns out that the other four coherent sheaves are given by
the following exact sequences that can be derived from the Euler sequence
(given for $\BP^n$ below though we only need the case of $n=4$ here)
associated with $\Omega^p(p) \equiv
\wedge^p \Omega \otimes {\cal O}(p)$
\begin{equation}
0\rightarrow \Omega^{p}(p)
\rightarrow {\cal O}^{\oplus ({}^{n+1}_{~p})}\rightarrow
\Omega^{p-1}(p-1)\otimes {\cal O}(1)\rightarrow 0
\label{generalisedeuler}
\end{equation}
Note the appearance of the binomial coefficients $({}^{n+1}_{~p})$ in
the above sequences.

The boundary fermion construction naturally leads to the spinor
bundle on $E$ rather than the coherent sheaf $E$. 
In the GLSM construction to obtain just the coherent sheaf
we restrict to one-particle states in the
corresponding boundary state. It was observed in \cite{bundlesglsm}
(see sec. 5.3) that when $E$ is the cotangent bundle, the spinor bundle
decomposes at different fermion numbers\footnote{For the case of weighted
projective spaces associated with one K\"ahler modulus Calabi-Yau manifolds, 
one replaces the fermion number by the $U(1)$ ($\BZ_K$) charge.}
to the different fractional branes. Thus, the monodromy about the LG point
is realised by  suitably changing the restriction on the fermion number of the
states. Thus, the five fractional branes for the quintic are in one-to-one
correspondence with the states: (the vacuum $|0\rangle$ satisfies
$\bar{\xi}_i |0\rangle =0$)
$$
|0\rangle\quad,\quad
\xi_i|0\rangle\quad,\quad
\xi_i\xi_j|0\rangle\quad,\quad
\xi_i \xi_j \xi_k |0\rangle\quad,\quad
\xi_i \xi_j \xi_k \xi_l |0\rangle
$$
subject to the condition $\phi_i \bar{\xi}^i=0$ being imposed. 
(A related observation was made by Mayr in 
\cite{mayr} where he referred to the fractional branes as providing a 
fermionic basis for branes). 

We define (using the notation in \cite{Ashok})
\begin{equation}
\boxed{
{\cal M}_i \equiv (-)^i \Omega_{\BP^4}^i(i)\Big|_{\textrm{quintic}}
\quad,\ i=0,1,2,3,4
}
\end{equation}
These branes ${\cal M}_i$ can be identified as the result of the analytic
continuation to large-volume of the
RS states in the Gepner model for the quintic. Under the quantum $\BZ_5$
symmetry(generated by $g$), one has
$$
g:\quad {\cal M}_i \rightarrow {\cal M}_{i+1\textrm{ mod }5}
$$

A basic test of this identification is based on the idea that we expect
the intersections of these branes (which is a topological quantity computed
by the open-string Witten index) to be the same at the orbifold point
and at large volume. At the orbifold point the intersections can be readily
computed from CFT techniques. For coherent sheaves on a smooth CY we
can compute the intersections using standard methods from differential
geometry. The two must agree and they indeed do so for the $L_i \equiv 0$
states from among the RS states in the Gepner model and the ${\cal M}_i$
that we have just described above.  

\subsection{Fractional two-branes from generalised Euler sequences}

We are now ready to discuss the case of fractional two-branes. 
As we have already seen, the Neumann boundary condition on the combination
$(\phi_1-\phi_2)$ is obtained from the supersymmetric variation of
the boundary condition $\xi_1=\xi_2$. Thus, on the boundary, given
these boundary conditions, we have effectively {\em four} independent
fermionic multiplets.  These fermionic multiplets are still subject
to the condition (\ref{gaugino}) imposed by the gaugino.
After eliminating $\bar{\xi}_2$ in favour of $\bar{\xi}_1$ using
Eq. (\ref{quinticbc}), 
Eq. (\ref{gaugino}) can be re-written as the following condition
\begin{equation}
\big(\phi_1 + \phi_2) \bar{\xi}_1 + \sum_{i=3}^5 \phi_i \bar{\xi}_i =0
\label{conda}
\end{equation}
Thus, this is equivalent to having four boundary fermions subject to the
one condition above. Unlike the case of fractional zero-branes, we see
that Eq. (\ref{conda}) is trivially satisfied when
$\phi_1+\phi_2=\phi_3=\phi_4=\phi_5=0$. This is possible on $\BP^4$,
when $\phi_1-\phi_2\neq 0$. Note that these conditions specify a 
two-brane (denoted below 
by $P$) in the manifold which is the  
resolution of the $\BC^5/\BZ_5$ singularity. It is trivial to see that
this two-brane restricts to a point
on the quintic. Away from the two-brane $P$,
Eq. (\ref{conda}) does reduce the number of fermions to three. 
This implies that the fermions are sections of the
sheaf $F_1$ given by the following sequence
\begin{equation}
0\rightarrow F_1 \rightarrow {\cal O}^{\oplus 4}
\rightarrow {\cal O}(1)\rightarrow {\cal X}_P\rightarrow 0
\end{equation}
The term involving ${\cal X}_P$ has been added to take care of
the fact that (\ref{conda}) is trivially satisfied on $P$. 
The following comments are in order here:
\begin{enumerate}
\item ${\cal X}_P$, by definition, vanishes away from $P$. In
particular, it vanishes on the $\BP^3\in \BP^4$ where $(\phi_1-\phi_2)=0$.
\item Defining 
$$
0\rightarrow F_0 \rightarrow {\cal O}\rightarrow {\cal X}_P\otimes {\cal O}(-1)
\rightarrow0\ ,
$$
the sequence can be rewritten as
\begin{equation}
0\rightarrow F_1 \rightarrow {\cal O}^{\oplus 4}
\rightarrow F_0\otimes {\cal O}(1)\rightarrow 0
\end{equation}
\item $F_0$ restricts to
${\cal O}_{\BP^3}$ on the $\BP^3$ not containing $P$.
\item The above sequence when restricted to the $\BP^3$ 
not containing $P$ becomes the
Euler sequence. Thus $F_1|_{\BP^3}=\Omega_{\BP^3}^1(1)$.
\item Note that $F_1$ is however a sheaf on $\BP^4$ even though the
sequence which generates it is reminiscent of the Euler sequence 
for $\BP^3$.
\end{enumerate}

The above identification suggests that the remaining fractional branes
can be given by exact sequences that on restriction 
to a $\BP^3$ not containing $P$ give the generalised Euler sequences 
of $\BP^3$. Explicitly, we can write
\begin{eqnarray}
0\rightarrow F_0 \rightarrow {\cal O} \rightarrow 
{\cal X}_P \otimes {\cal O}(-1) \rightarrow 0 \nonumber \\
0\rightarrow F_1 \rightarrow {\cal O}^{\oplus 4} \rightarrow 
F_0 \otimes {\cal O}(1) \rightarrow 0 \nonumber \\
0\rightarrow F_2 \rightarrow {\cal O}^{\oplus 6} \rightarrow 
F_1 \otimes {\cal O}(1) \rightarrow 0 \label{neweuler}\\
0\rightarrow F_3 \rightarrow {\cal O}^{\oplus 4} \rightarrow 
F_2 \otimes {\cal O}(1) \rightarrow 0 \nonumber \\
0\rightarrow F_4 \rightarrow {\cal O} \rightarrow 
F_3 \otimes {\cal O}(1) \rightarrow 0\nonumber  
\end{eqnarray}
The last sequence which generates $F_4$ is rather interesting. 
One can argue that 
$F_4 = -{\cal X}_P\otimes {\cal O}(3)$. 
That $F_4$ must at least have ${\cal X}_P$ as a factor is clear
since the last sequence must restrict to zero on
on the $\BP^3\in \BP^4$ not containing 
$P$. This is because there is no corresponding generalized Euler sequence
that appears on $\BP^3$. The factor of ${\cal O}(3)$ can be deduced from
the general pattern that we observe in these sequences. 
We refer to $F_4$ as the {\em Coulomb
branch brane} because of this vanishing property 
on restriction to the $\BP^3$ not
containing $P$.  The remaining branes $(F_0,\ldots,F_3)$ 
will be called
as the {\em Higgs branch
branes}. As will be explained later, this parallels the {\em missing
branes} that one needed to make a correspondence (called the quantum
McKay correspondence) between D-branes on
a non-supersymmetric orbifold and D-branes on the Hirzebruch-Jung
resolution as considered by Martinec and Moore\cite{MM}. This relationship
will be made more precise in a later section.

The main claim we wish to make is that the new fractional branes of \cite{Ashok}
are to be identified with (a minus sign indicates an anti-brane)
\begin{equation}
\boxed{
{\cal F}_i = (-)^i F_i\Big|_{\textrm{quintic}}\  , \quad i=0,1,2,3,4
}
\end{equation}
provided we choose 
\begin{equation}
\textrm{ch}\ \big( {\cal X}_P\big)\big|_{\textrm{quintic}} = J^3/5 \ .
\end{equation}
where $J$ generates the K\"ahler class on the quintic and is normalised
such that $\langle J^3 \rangle_{\textrm{quintic}} = 5 \langle J^4
\rangle_{\BP^4}=5$. As a first check, we have verified that the
Chern classes of the ${\cal F}_i$ agree with those given by 
Ashok et. al. in \cite{Ashok}. (More details are provided in
appendix \ref{quinticintersection}.) We also propose that under the 
quantum $\BZ_5$ symmetry,
one has
$$
g:\quad {\cal F}_i \rightarrow {\cal F}_{i+1\textrm{ mod }5}
$$

We will motivate here this unusual assignment of Chern character for the 
object ${\cal X}_P$ while a more detailed justification of the appearance
of the $1/5$ factor will be provided in a later section.
It is clear from a simple argument that  
sheaves in the ambient projective space (obtained from the fractional 
branes by blowing up the orbifold singularity) 
when restricted to the Calabi-Yau would 
fail to give objects that have the charge of a
single zero-brane on the CY. A two-brane wrapping a $\BP^1\in\BP^4$ (which
intersects the quintic on a point) will have
Chern character $J^3+a J^4$ for some $a$. On restricting this
to the quintic, we obtain an object with Chern character $J^3$ which
has the charge of $5$ zero-branes. Thus if we need to produce an object
with the charge of a single zero-brane on the CY it appears that we must
begin with a sheaf on $\BP^4$ whose Chern character has leads off with
a $J^3/5$ term. However more is clearly needed to justify this unusual
choice. 
                                                                                
Note that with this assignment of fractional Chern characters to 
the object ${\cal X}_P$ the maps in the exact sequences for the 
$F_i$ that we have given no longer have any obvious and 
rigorous mathematical meaning. This is in contrast to the case of 
the large-volume analogs of the fractional zero-branes that we discussed
earlier where this a clear correspondence between physical constructions
in the GLSM and rigourous mathematical constructions. 
Nevertheless we will continue to use these exact sequences for the 
$F_i$'s at least as a convenient device or a mnemonic to write down
their Chern characters and hence their D-brane charges. 

The five new fractional two branes for the quintic are in one-to-one
correspondence with the states: (the vacuum $|0\rangle$ satisfies
$\bar{\xi}_i |0\rangle =0$ and $i=1,3,4,5$ below)
$$
F_0 \sim |0\rangle\ ,\quad
F_1\sim\xi_i|0\rangle\ ,\quad
F_2\sim\xi_i\xi_j|0\rangle\ ,\quad
F_3\sim\xi_i \xi_j \xi_k |0\rangle\ ,\quad
F_4\sim\xi_1 \xi_3 \xi_4 \xi_5 |0\rangle
$$
subject to the modified gaugino constraint in Eq. (\ref{conda}) being
imposed on them.

We now proceed to consider the case of fractional two-branes on
$\BC^n/\BZ_N$ as a somewhat simpler version of the quintic
example that we just considered. While the orbifold aspects are carried
in some generality, we will consider the large volume aspects in great
detail for $\BC^3/\BZ_3$ postponing the discussion to a future
publication.

\section{Fractional $2p$-branes in $\BC^n/\BZ_N$}

In this section, we will discuss the case of fractional $2p$-branes in
$\BC^3/\BZ_3$. We will do so from two perspectives: (i) the boundary states 
at the orbifold point and (ii) sheaves on the resolved space using the GLSM. 
We will first compute at the orbifold end two sets of intersection
numbers, the intersection numbers between fractional
two-branes and the fractional zero-branes and also the intersection 
numbers between the fractional two-branes themselves. 
Even at the orbifold point, though the computations are
well-known and we will use some of those results, 
we will emphasize some non-trivial features of the 
calculation that have not attracted due attention earlier.   
Fractional $2p$-branes on orbifolds have been considered, for instance
in refs. \cite{diacgom,Gaberdiel:1999ch,BCR,tadashi,BDFLM,romel1}.

As we indicated earlier we expect that the  
intersection numbers that we obtain at the orbifold end
will be reproduced by the sheaves that correspond
to these objects when
the K\"ahler modulus is continued to the large-volume region of 
the K\"ahler moduli space.  
We will see that identifying at the large-volume point the sheaves 
corresponding to the fractional two-branes is considerable more involved
than in the case of the zero-branes. However, we will show that 
the objects that we identify at large volume reproduce precisely 
the intersection numbers that were computed by CFT methods at the 
orbifold point. 
However, we will include no discussion on their stability and thus
will be focusing on the topological B-branes.

We choose the orbifold action given by
$$
\phi_i \longrightarrow e^{2\pi \nu_i}\ \phi_i \ ,
$$
which we will compactly write as
$[\nu_1,\nu_2,\ldots,\nu_n]\equiv\frac1N[a_1,a_2,\ldots,a_n]$  for some integers
$a_i$.
Further, the type II GSO projection will require
us to choose $\sum_i \nu_i =0$ mod $2$ (rather than mod $1$).
In fact, we will require something a little bit more stringent in 
the sequel.  The boundary states that
we construct are similar in spirit to the ones constructed in
ref. \cite{gaberdiel} (see in particular section 4.1 for the
discussion on the GSO projection) for a single ${\cal N}=2$ chiral multiplet.
We will however not get into the details of the GSO
projection because we do not include the spacetime part. They can
be included in a straightforward fashion.

\subsection{Fractional zero-branes in $\BC^n/\BZ_N$}

The case of fractional zero-branes has been discussed in great detail in
the paper by Diaconescu and Gomis\cite{diacgom}. 
Since the 
orbifold action in the open-string sector is straightforward, 
the only difference between the case of Neumann boundary conditions
and Dirichlet boundary conditions arises from the 
zero-mode sector. In the non-zero mode sector the open-string partition
function is identical. 

We first write down the partition function in the zero-brane case.
In the open-string channel, the amplitude in the $m$-th twisted sector is
given by \cite{diacgom} (the notation is as in \cite{diacgom} as well)
\begin{eqnarray}
{\cal A}^{(0)}_m &=& {\rm tr}\Big( g^m\ \frac{1+(-)^F}2\ e^{-2tH_o} \ \Big)\nonumber \\
&=&
 V_1 \int_0^\infty \frac{dt}{2t} (8\pi^2 \alpha' t)^{-1/2}
\times
\prod_{j=1}^n \Big[ \frac{2\sin(\pi m \nu_j) \eta(it)}{\theta_1(m\nu_j,it)}\Big]
\\
&&\times
\frac12 \Big[
\prod_{j=1}^n\Big(\frac{\theta_3(m\nu_j,it)}{\eta(it)}\Big)
\prod_{j=1}^n\Big(\frac{\theta_4(m\nu_j,it)}{\eta(it)}\Big)
\prod_{j=1}^n\Big(\frac{\theta_2(m\nu_j,it)}{\eta(it)}\Big)
\Big]\nonumber
\end{eqnarray}
In the above expression, the second line is the contribution from the 
worldsheet bosons and the third line is the contribution from the worldsheet
fermions.
In the second line, $V_1 (8\pi^2 \alpha' t)^{-1/2}$ is from the bosonic
zero-modes and the other term is from the non-zero modes (see Eq. (B.6)
of \cite{BCR}, for instance).
Note that if either $m=0$ or some particular $\nu_i=0$, then we need to use
the following identity:
$$
\lim_{\nu\rightarrow0}
\Big[\frac{2\sin(\pi \nu) \eta(\tau)}{\theta_1(\nu,\tau)}\Big]
=\frac{1}{\eta^2(\tau)}\ .
$$

To go to the closed-string channel, we consider the modular transform
of the above amplitude,i.e., $\tau=it \rightarrow -1/\tau\equiv 2il$.
\begin{eqnarray}
{\cal C}^{(0)}_m &=&
 V_1 (8\pi^2 \alpha' )^{-1/2}\int_0^\infty \frac{dl}{2l}
\times  l^{1/2} \ \Big[\frac{1}{2l\ \eta^2(2il) }\Big]^r \times
\prod_{j=1}^{n-r} \Big[
\frac{(-i) 2\sin(\pi m \nu_j) \eta(2il)}{\theta_1(-2il m\nu_j,2il)}\Big]
\\
&\times&
\frac12 \Big[
\prod_{j=1}^n\Big(\frac{\theta_3(-2il m\nu_j,2il)}{\eta(2il)}\Big)
\prod_{j=1}^n\Big(\frac{\theta_4(-2il m\nu_j,2il)}{\eta(2il)}\Big)
\prod_{j=1}^n\Big(\frac{\theta_2(-2il m\nu_j,2il)}{\eta(2il)}\Big)
\Big]\nonumber
\end{eqnarray}
In the above expression, $r$ is the number of directions for which
$m\nu_i=0$. Thus, when $m=0$, one has $r=n$ and when $m\neq0$, then
$r$ is the number of directions on which the orbifolding group has no
action.
One looks for a (GSO projected) state $|\!|B0,m\rangle\!\rangle$ in
the $m$-th twisted sector for
which\footnote{This is not quite the boundary state that satisfies
Cardy's condition and hence we represent it by
$|\!|B0,m\rangle\!\rangle$ rather than $|B0,m\rangle$ to avoid
confusion.}
\begin{equation}
{\cal C}^{(0)}_m = \int_0^\infty dl\ \langle\!\langle B0,m|\!|\ e^{-l H_c}\
|\!|B0,m\rangle\!\rangle
\end{equation}

Since we will not need much detail, the dedicated reader may
obtain the precise form of the $|\!|B0,m\rangle\!\rangle$ 
from Eqs. (4.14-4.23) of \cite{diacgom} except for a small
difference. We remove the character (for the irrep I of $\BZ_N$)
$\chi_I(g^m)$ (as given in Eq. 4.23 of \cite{diacgom})
since  we wish to include as a part of the normalisation.   
It is useful to note 
that $|\!|B0,m=0\rangle\!\rangle$ is the boundary state for a zero-brane in
flat-space. The consistent boundary states are labelled by the irreps of
$\BZ_N$ (satisfying Cardy's
consistency conditions) for the fractional zero branes are
\begin{equation}
\boxed{
|B0:I\rangle = \sum_{m=0}^{N-1} \psi^{(0)~m}_I\ |\!|B0,m\rangle\!\rangle\quad
I=0,1,\ldots, (N-1)
\label{zerobs}
}
\end{equation}
where $\psi^{(0)~m}_I=\chi_I(g^m)/
\sqrt{N}= e^{2\pi i Im/N}/\sqrt{N}$ 
is the normalisation for the fractional zero-branes. The D0-brane charge
comes from the RR charges in the untwisted sector and is $1/N$ the value
in flat space -- a $1/\sqrt{N}$ from the normalisation $\psi^{(0)~0}_I$ and
another $1/\sqrt{N}$ from the ``renormalisation'' of the charge in the
orbifolded space\cite{BCR}. The RR charge from the m-th twisted sector
is
\begin{equation}
{Q^{(0)~m}_I} = \frac{\chi_I(g^m)}{N} 
\end{equation}

\subsection{Fractional $2p$-branes}

We will now consider the case where one imposes Neumann boundary conditions
on one of the fields on which the orbifold group has a non-trivial action.
In computing the open-string amplitude, the bosonic non-zero contribution
and the fermionic ones are unchanged. However, one has to treat the bosonic
zero-modes separately due to the  momentum zero mode.  The contribution
from the bosonic zero-modes to the open-string partition function with
a $g^m$ insertion is given by (for a fractional two-brane):
\begin{equation}
\textrm{bosonic zero-mode contribution} = \left\{\begin{matrix}
V_3 (8\pi^2 \alpha' t)^{-3/2} & m=0 \\[5pt]
V_1  (8\pi^2 \alpha' t)^{-1/2}(4\sin^2 \pi m \nu)^{-1}
& m\neq0
\end{matrix}\right.
\end{equation}
The $m=0$ case is the same as the one with no orbifolding.
The $m\neq0$ sectors are similar to the zero-brane case that we just
considered except for an {\em additional factor} of $(4\sin^2 \pi m \nu)^{-1}$.
                                                                                
Putting all this together, one obtains the following changes in the
expressions for ${\cal A}_m^{(2p)}$ (for fractional $2p$-branes,
the index $a=1,\ldots,p$ runs over the Neumann directions)
with respect to the zero-brane case
given earlier, i.e., ${\cal A}_m$.
\begin{eqnarray}
V_1 (8\pi \alpha't)^{-1/2} \longrightarrow V_{2p+1}
(8\pi \alpha't)^{-(2p+1)/2} \quad &m=0 \nonumber \\
V_1 \longrightarrow V_1 \prod_{a=1}^{p} (4\sin^2 \pi m \nu_a)^{-1} \qquad\qquad& m\neq0
\nonumber
\end{eqnarray}
The annulus amplitude for the $2p$-brane in the untwisted ($m=0$) sector
is thus
\begin{eqnarray}
{\cal A}^{(2p)}_0 &=& {\rm tr}\Big( g^m\ \frac{1+(-)^F}2\ e^{-2tH_o} \
\Big)\nonumber \\
&=&
 V_{2p+1} \int_0^\infty \frac{dt}{2t} (8\pi^2 \alpha' t)^{-(2p+1)/2}
\times
\prod_{j=1}^n \Big[ \frac{2\sin(\pi m \nu_j)
\eta(it)}{\theta_1(m\nu_j,it)}\Big]
\\
&&\times
\frac12 \Big[
\prod_{j=1}^n\Big(\frac{\theta_3(m\nu_j,it)}{\eta(it)}\Big)
\prod_{j=1}^n\Big(\frac{\theta_4(m\nu_j,it)}{\eta(it)}\Big)
\prod_{j=1}^n\Big(\frac{\theta_2(m\nu_j,it)}{\eta(it)}\Big)
\Big]\nonumber
\end{eqnarray}
and in the $m\neq0$ sectors
\begin{eqnarray}
{\cal A}^{(2p)}_m &=& {\rm tr}\Big( g^m\ \frac{1+(-)^F}2\ e^{-2tH_o} \
\Big)\nonumber \\
&=&
 V_1 \prod_{a=1}^{p} (4\sin^2 \pi m \nu_a)^{-1}
\int_0^\infty \frac{dt}{2t} (8\pi^2 \alpha' t)^{-1/2}
\times
\prod_{j=1}^n \Big[ \frac{2\sin(\pi m \nu_j)
\eta(it)}{\theta_1(m\nu_j,it)}\Big]
\\
&&\times
\frac12 \Big[
\prod_{j=1}^n\Big(\frac{\theta_3(m\nu_j,it)}{\eta(it)}\Big)
\prod_{j=1}^n\Big(\frac{\theta_4(m\nu_j,it)}{\eta(it)}\Big)
\prod_{j=1}^n\Big(\frac{\theta_2(m\nu_j,it)}{\eta(it)}\Big)
\Big]\nonumber
\end{eqnarray}

The boundary state is quite similar to the one for the fractional
zero-branes as given in Eq. (\ref{zerobs}) with the following
replacements:
\begin{eqnarray}
|\!|B2p:m=0\rangle\!\rangle &\equiv& |B2p\rangle_{\textrm{flat space}}\\
|\!|B2p:m\neq0\rangle\!\rangle &\equiv& 
\widetilde{|\!|B0,m\rangle\!\rangle}\ ,
\end{eqnarray}
where the tilde represents the operation which switches the signs on the
non-zero modes in a manner suitable for a $2p$-brane. With this, we
can write the boundary state for the fractional $2p$-branes:
\begin{equation}
\boxed{
|B2p:I\rangle = \sum_{m=2p}^N \psi^{(0)~m}_I\
|\!|B2p,m\rangle\!\rangle\quad
I=0,1,\ldots, (N-1)
\label{2pbs}
}
\end{equation}
where 
$$
\psi^{(2p)~m}_I=\frac{\chi_{I}(g^m)}
{\sqrt{N}\ \prod_{a=1}^{p} (-2i\sin\pi m\nu_a)}= \frac{e^{2\pi i
Im/N}}{\sqrt{N}\prod_{a=1}^{p} (-2i\sin\pi m\nu_a)}\ ,
$$
where we have included a constant phase factor of $(-i)$ along with the 
$2\sin\pi m\nu_a$ since it makes  all intersection numbers being real.
The above normalisation implies that the $m$-th twisted sector part of the
boundary state for a fractional two-brane
will be the same as the fractional branes with a multiplicative factor
of $(-2i \sin \pi m \nu)^{-1}$ (for every Neumann boundary condition)
and thus the RR-charge in the $m$-th
twisted sector  of the two-brane is given by
\begin{equation}
Q^{(2p)~m}_I = \frac{\chi_I(g^m)}{N} \prod_{a=1}^{p} \frac1{2 \sin \pi m
\nu_a}
\end{equation}
where the index $a$ runs over the Neumann directions and $I=1,\ldots,N$
label the fractional $2p$-branes. Finally, the
two-branes all carry $2p$-brane RR charge from the untwisted sector
which is $1/N$ of the result in flat space.

\subsection{Intersection numbers}

From the appendix of \cite{BCR}, 
one can see that computation of the
open-string Witten index is given by (Eq.. (4.54) in \cite{BCR})
\begin{equation}
{\cal I}^{p-p'}_{IJ} =  \sum_{m\neq0}  \big(\psi^{(p)~m}_I\big)^* \psi_J^{(p')~m} \prod_{i=1}^n (-2i\sin \pi m\nu_i)
\end{equation}
where $\psi_I^m$ are the normalisations associated with the boundary condition
$I$. For the fractional zero-branes, one has
$$
\psi_I^{(0)~m} = \frac{1}{\sqrt{N}} \ \chi_I(g^m) \ ,
$$
where $\chi_I(g^m)=\exp(2\pi i mI)$  for $\BZ_n$. Note that one drops out
the spacetime contribution to this since it multiplies the above result by
zero.

We can also work out the general formulae for the 
open-string Witten index for open-strings
that connect fractional $2p$-branes to fractional $2p'$-branes.
This generalises the expression existent in the literature for the
case of fractional zero-branes. A straightforward computation gives the
following {\em master formula} for the $\BC^n/\BZ_N$
orbifold\cite{BCR,romel1}:
\begin{equation}
\boxed{
{\cal I}^{2p.2p'}_{I,J} = -\frac{(-i)^{n+p-p'}}{N} \sum_{j=1}^{N}\!{}^{'}
e^{2\pi i (I-J)j/N} \times
\frac{\prod_{k=1}^{n}(2\sin\pi j \nu_k)}{\prod_{a=1}^{p} (2 \sin \pi j
\nu_a) \prod_{b=1}^{p'} (2 \sin \pi j
\nu_b)}
\label{master}
}
\end{equation}
where the product in the denominator of the RHS runs over the $p$ ($p'$)
Neumann directions alone and the prime indicates that the sum does
{\em not} include terms that have vanishing denominators -- this happens
when $j\nu_a$ become integers. One can see that when the 
intersection between fractional $2p$-branes and fractional $2(n-p)$
branes [obtained by exchanging Neumann and Dirichlet boundary
conditions] is the identity matrix. 
   
\subsection{Intersection numbers -- examples}

\subsubsection{$\BC^3/\BZ_3$}

The $\BZ_3$ action is taken to be $\frac13[-2,-2,-2]$ with the Neumann
boundary condition
chosen on the first field when fractional two-branes are
considered\footnote{If we choose the Neumann direction
to have $\nu=1$ rather than $\nu=-2$, the intersection numbers are
non-integral. This is related to the type II open-string GSO
projection\cite{gaberdiel}.
}.
There are three fractional zero and two-branes, which we will represent by
$S_I^{0}$ and $S_I^{2}$ respectively. The quantum $\BZ_3$ acts on these 
branes by shifting $I\rightarrow I+1\textrm{ mod }3$. We will write the
intersection numbers in terms of the generator $g$ of the $\BZ_3$.

The master formula, Eq. (\ref{master})
gives on using $2\sin(\pi/3)=2\sin(2\pi/3)=\sqrt{3}$
\begin{eqnarray}
{\cal I}^{0,0} &=& -(1-g)^3 \nonumber \\
{\cal I}^{0,2} &=& -g^2(1-g)^2 \\
{\cal I}^{2,2} &=& g(1-g) \nonumber 
\end{eqnarray}                                  
Note that in the expression for the 
intersection form ${\cal I}^{0,2}$ between fractional zero
and two-branes, the factor of $g^2$ can be gotten rid of by
relabelling/shifting the $I$ labels, of say, the fractional two-branes
by two. Note that such a shift does not affect ${\cal I}^{2,2}$. This
has to be kept in mind while comparing with the intersection form
for the coherent sheaves that we propose as candidates for the
large-volume analogs of the fractional two-branes in the next
subsection.

\subsubsection{$\BC^5/\BZ_5$}

The $\BZ_5$ action is taken to be $\frac15[-4,-4,-4,-4,-4]$. 
We will consider the cases of fractional zero, two and four branes.
Again, we will use the notation $S_I^{(2p)}$ with
$I=1,2,3,4,5\textrm{ mod }5$ and the quantum $\BZ_5$ being generated by
$g$ which takes $S_I^{(2p)}\rightarrow S_{I+1}^{(2p)}$. The various
intersection matrices are given by:
\begin{eqnarray}
{\cal I}^{0,0} &=& -(1-g)^5 \nonumber \\
{\cal I}^{0,2} &=& -g^3(1-g)^4 \nonumber \\
{\cal I}^{0,4} &=& -g(1-g)^3 \\
{\cal I}^{2,2} &=& g(1-g)^3 \nonumber \\
{\cal I}^{2,4} &=& g^4(1-g)^2 \nonumber \\
{\cal I}^{4,4} &=& -g^2(1-g) \nonumber 
\end{eqnarray}     

\subsection{Coherent sheaves on the resolved space}
\subsubsection{The basic idea for fractional branes}

\noindent
We will now discuss in greater detail the nature of the coherent
sheaves that arise from the continuation to large-volume of the fractional 
p-branes that we have been studying at the orbifold end. 
For specificity we will focus on the case of fractional
2-branes in the case of the blow-up of the orbifold $C^3/Z_3$. In this
case the manifold at large volume is the total space of the line 
bundle ${\cal O}(-3)$ on $\BP^2$. 

We summarise in the table below how some simple examples of branes 
in this manifold can be represented by coherent 
sheaves, or equivalently the corresponding sequences. All
these examples can be produced from the fractional zero-branes continued
to large volume or by considering bound states of these objects.     
\begin{center}
\begin{tabular}{l|c|c}\hline
\hspace{0.8in}Object & the associated sheaf  & Chern ch. \\ \hline
A 4-brane wrapping $\BP^2$ &
${\cal O}_{\BP^2}$ 
& $1$ \\
A 2-brane wrapping a $\BP^1\subset \BP^2$ &
${\cal O}_H\equiv [{\cal O}_{\BP^2}(-1)\rightarrow {\cal O}_{\BP^2}]$ 
& $J -\frac{J^2}2$ \\
A point on $\BP^2$ 
by & ${\cal O}_{\rm pt}\equiv [{\cal O}_{\BP^2}(-1)\rightarrow
{\cal O}^{\oplus 2}_{\BP^2}\rightarrow {\cal O}_{\BP^2}(1)]$
& $J^2$ \\ \hline
\end{tabular}
\end{center}
where $J$ generates $H^2(\BP^2,\BZ)$ and $\langle J^3\rangle_{\BP^2}=1$.
Note that we can always twist the 2-brane by tensoring it with ${\cal
O}_{\BP^2}(n)$. This changes the $J^2$ part in the Chern character.

We shall now consider the basic two-cycles in this non-compact Calabi-Yau.
$M\equiv {\cal O}_{\BP^2}(-3)$, which is the crepant resolution of the 
$\BC^3/\BZ_3$ orbifold that we just considered: 
\begin{itemize}
\item
There is one
{\em compact} two-cycle given by a $\BP^1$ in $\BP^2$. 
\item In addition, there are
three {\em non-compact} two-cycles corresponding to the fibre over $\BP^2$. 
These two-cycles
intersect the boundary at infinity, which is a $S^5/\BZ_3$, on a
one-cycle $\gamma\in S^5/\BZ_3$. ($\gamma$ is a element of
$H_1(S^5/\BZ_3,\BZ)=\BZ_3$.)
\end{itemize}
\begin{figure}
\centering
\includegraphics[height=2.3in]{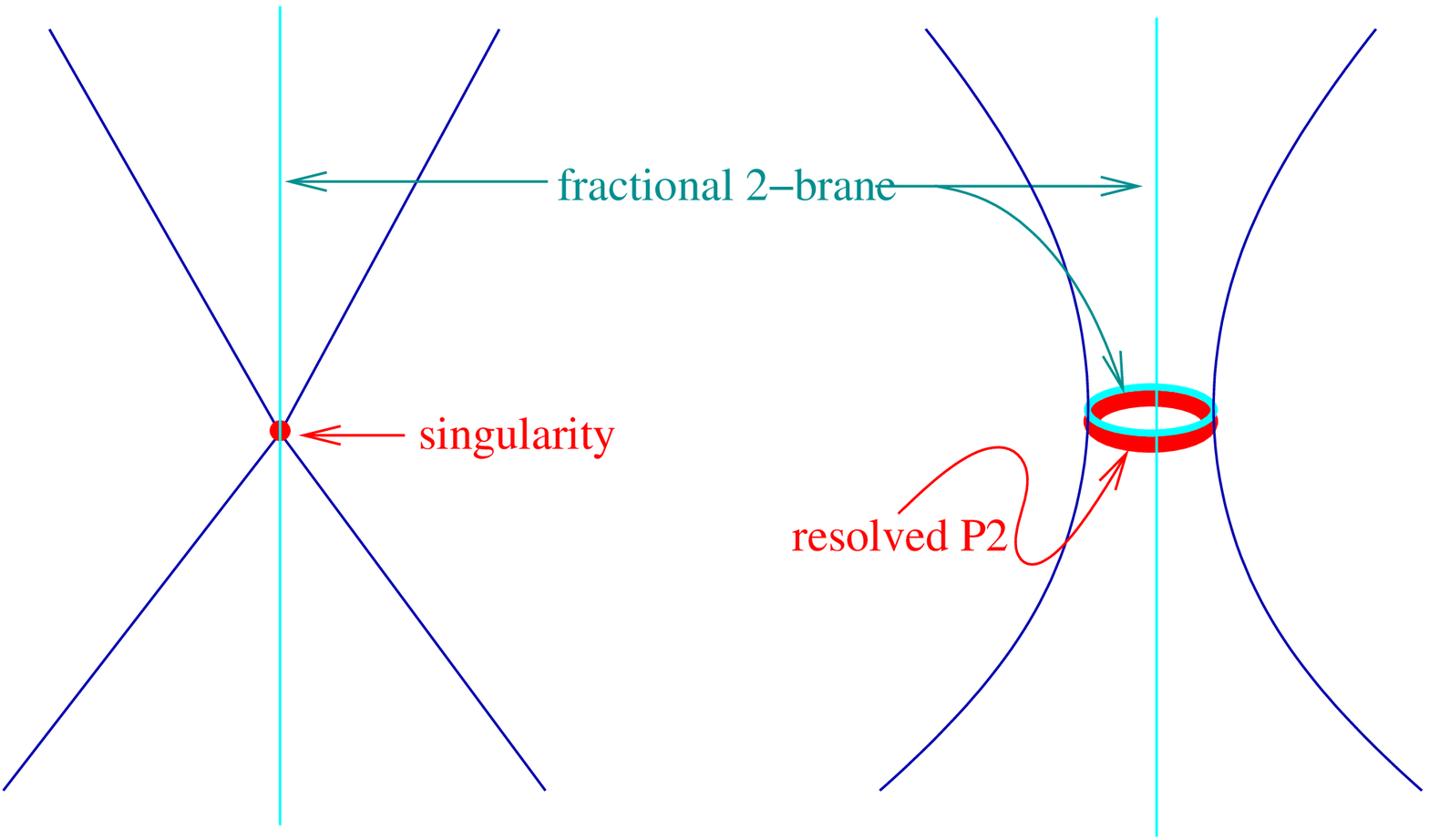}
\caption{A schematic description of the fractional two-branes both
before and after the singularity is resolved.}
\end{figure}
Consider a two-brane wrapping a non-compact two-cycle.
What is the the representation for such
branes? In particular, what is its D-brane charge as given by 
some appropriate Chern character? 

In the case of non-compact manifolds the correct framework in which 
to discuss D-brane charges is compact cohomology or equivalently 
relative cohomology.
For a non-compact manifold $M$ with boundary $N$, 
the two-brane charges take values in $H^2_{\textrm{compact}}(M,\BZ)\sim 
H^2(M,N,\BZ)$. A calculation (given in the appendix \ref{cohomology}) 
shows that
this is $\BZ$ and since $M$ has only one compact cycle, i.e., the
$\BP^2$, the basic two-brane is obtained by wrapping a $\BP^1\in\BP^2$.
Hence, $J$ generates $H^2(M,N,\BZ)$. However objects that wrap the
non-compact two-cycle of $M$ will have a charge in $H^2(M,\BZ)\sim\BZ$.
The two cohomologies are related by a long exact sequence, the relevant
part of which for
our case reduces to the following (see the appendix  \ref{cohomology}
for details)
$$
0\rightarrow H^{2}({\cal O}_{\BP^2}(-3), S^5/\BZ_3;\BZ) \rightarrow 
H^{2}({\cal O}_{\BP^2}(-3);\BZ)
\stackrel{j}{\rightarrow} H^2(S^5/\BZ_3;\BZ) 
 \rightarrow 0 
$$
\begin{equation}
0 \rightarrow \BZ \stackrel{3}{\rightarrow} \BZ \rightarrow \BZ_3 \rightarrow 0
\end{equation}
Let $J'$ generate $H^{2}({\cal O}_{\BP^2}(-3);\BZ)$. The exact sequence above
indicates that $[J]\sim 3[J']$. Thus if $[J]$ is the charge basis of the
2-brane in the compact cohomology, then a charge $+1$ 2-brane in 
the compact cohomology would correspond to a charge $+3$ 2-brane
in the $[J']$ basis.
A single 2-brane wrapping the non-compact two-cycle would have charge
$+1$ in the $[J']$ basis or charge $1/3$ in the $[J]$ basis. Thus  
three two-branes wrapping a non-compact two-cycle of $M$ can give you
an object which is equivalent\footnote {The term equivalent can be made 
precise in the toric description of this geometry, where the equivalence
is a consequence of the linear equivalence of divisors.}
to a two-brane in  $\BP^2$ as elements of 
$H^{2}({\cal
O}_{\BP^2}(-3), S^5/\BZ_3;\BZ)$. In many ways, this is like the
fractional zero-branes -- the fractional
zero-branes were localised at the
singularity and couldn't be moved away from there unless three of them
were taken to form a regular zero-brane. 

Thus we have motivated the existence of objects with fractional two-brane
charge measured in the charge basis associated with the compact cohomology.
We can also perform an equivalent computation in the context of K-theory
rather than cohomology, showing again the existence of fractional 
two-brane charges, but now obviously in the compact K-group associated 
with the $\BP^2$. This computation not being essential at this point, is 
relegated to an appendix\ref{ktheory}. 

We can now write down the form of the Chern character for such a fractional 
two-brane. Clearly its leading term must be of the for $J/3$, while the 
$J^2$ term may depend on whether we have twisted the object by a line
bundle ${\cal O}(n)$. Thus the general form will be 
\begin{eqnarray}
\ch({\cal X}_m) &=& \frac1{3} \textrm{ch}[{\cal
O}_{\BP^2}(m-1)\rightarrow {\cal O}_{\BP^2}(m)] \nonumber \\
&=& \frac{1}{3}\Big(J + \frac{(2m-1)}2J^2 \Big)
\end{eqnarray}
where by construction,  ${\cal X}_m$ carries $1/3$ the charge of a two-brane on
$\BP^2$ after including a twist which we indicate by the subscript $m$.
Note that ${\cal X}_{m+1} = {\cal X}_m\otimes
{\cal O}_{\BP^2}(1)$ and that $\textrm{ch}({\cal X}_{m+1}) =
\textrm{ch}({\cal X}_m)
+\frac13\ \textrm{ch}({\cal O}_{\rm pt})$.

\subsubsection{Fractional two-branes for  $\BC^3/\BZ_3$}

Let us consider the case of fractional two-branes in the $\BC^3/\BZ_3$
example. Let us impose Neumann boundary conditions on $(\phi_1-\phi_2)$ and
Dirichlet on $(\phi_1+\phi_2)$ and $\phi_3$\footnote{We choose this
boundary condition since this will be compatible to adding a
superpotential $G(\phi)=\phi_1^3 + \phi_2^3 + \phi_3^3$.} . 
Away from  the orbifold point, 
thus there are two fermions (after eliminating $\bar{\xi}_2$ in favour
of $\bar{\xi}_1$): $\bar{\xi}_1$ and $\bar{\xi}_3$. 
The gaugino constraint is
\begin{equation}
(\phi_1+\phi_2) \bar{\xi}_1 + \phi_3 \bar{\xi}_3  =0\ .
\end{equation}
Thus, when $(\phi_1+\phi_2,\phi_3)\neq (0,0)$, the constraint removes
one fermion and 
when $(\phi_1+\phi_2,\phi_3) = (0,0)$ the constraint is trivially satisfied.
(This  is possible, when  $\phi_1-\phi_2\neq 0$.)
Repeating the arguments for $\BC^5/\BZ_5$, we get three fractional
two-branes given by the sequences 
\begin{eqnarray}
0\rightarrow F_0 \rightarrow {\cal O}_{\BP^2} \rightarrow
{\cal X}_{n-1} \otimes {\cal O}_{\BP^2}(1) \rightarrow 0 \\
0\rightarrow F_1 \rightarrow {\cal O}_{\BP^2}^{\oplus 2} \rightarrow
F_0 \otimes {\cal O}_{\BP^2}(1) \rightarrow 0 \\
0\rightarrow F_2 \rightarrow {\cal O}_{\BP^2} \rightarrow
F_1 \otimes {\cal O}_{\BP^2}(1) \rightarrow 0 
\end{eqnarray}
In the first line, we have included the fractional contribution 
by inserting ${\cal X}_{n-1}$ in the first line of the above
equation to complete the sequence.

We then obtain the following identifications
\begin{eqnarray}
\ch(F_0) &=& \ch\big[{\cal O}_{\BP^2}\big] - \ch({\cal X}_{n-1}) \\
-\ch(F_1) &=& \ch\big[{\cal O}_{\BP^2}(1)-{\cal O}_{\BP^2}^{\oplus 2}\big] 
- \ch({\cal X}_{n}) \\
\ch(F_2) &=& \ch\big[{\cal O}_{\BP^2}-{\cal O}_{\BP^2}^{\oplus 2}(1)
+{\cal O}_{\BP^2}(2)\big] 
- \ch({\cal X}_{n+1}) 
\end{eqnarray}
The objects in the square brackets  are non-fractional objects and hence
correspond to coherent sheaves on $\BP^2$. Thus these terms must
necessarily arise from the twisted sectors of the boundary state.
The contribution of the untwisted sector is contained in 
the term containing the ${\cal X}$'s. 

Now the Chern character add up as follows:
$$
\textrm{ch}(F_0 - F_1 + F_2) = \ch\big[{\cal O}(2) - {\cal
O}(1) \big] -J + \frac{(2n+1)}2 J^2
$$
where we have kept the two contributions separate. Note that the ${\cal
X}$'s
have summed up to give an object that has the 
Chern class of a two-brane on $\BP^2$.

The Euler form associated with these fractional two-branes,
$F_i$ have been computed in the appendix \ref{c3intersections}
and are generically fractional.
However, the intersection form which is obtained by antisymmetrisation
of the Euler form is integral\cite{dfr1}. The integrality of the
intersection form implies that the charge quantisation condition
is  not violated. After restricting these sheaves to the Fermat
cubic hypersurface in $\BP^2$, the Euler form reduces to the
same intersection matrix without any need for antisymmetrisation.
Finally, up to trivial shifts, the intersection matrices also agree
with the open-string Witten index computed at the orbifold end. This
provides further evidence towards the identification of the $(-)^IF_I$ 
as the analytic continuation of
the fractional two-branes $S^{(2)}_I$ that we constructed at the orbifold
end.

A more precise mathematical statement is to
write the $F_i$ as sheafs in the total space. Let $i$ be the inclusion
of the $\BP^2$ in the total space ${\cal O}_{\BP^2}(-3)$ and $i_*[E]$
represent the push-forward of the bundle $E$ on $\BP^2$ to the space
${\cal O}_{\BP^2}(-3)$
\begin{eqnarray}
F_0 &\sim& i_*\big[{\cal O}_{\BP^2}\big] - {\cal X}_{n-1} \\
F_1 &\sim& i_*\big[
{\cal O}_{\BP^2}^{\oplus 2}\rightarrow {\cal O}_{\BP^2}(1)\big] 
- {\cal X}_{n} \\
F_2 &\sim& i_*\big[{\cal O}_{\BP^2}\rightarrow {\cal O}_{\BP^2}^{\oplus 2}(1)
\rightarrow {\cal O}_{\BP^2}(2)\big] 
-{\cal X}_{n+1}
\end{eqnarray}
Chern classes on the non-compact space can include terms associated with
non-compact divisors. In particular, a term such as $D_1\cdot D_2$ can
appear. Indeed, one has
$\ch({\cal X}_n) = D_1\cdot D_2 + \cdots$, where $D_1$(resp. $D_2$) is the 
divisor associated to $\phi_1=0$ (resp. $\phi_2=0$) and the ellipsis
contains terms associated with the Chern class of a point.
Note that there is nothing fractional
about ${\cal X}_n$ here. The intersection numbers for the above
sheafs can be computed directly in the non-compact space and it
reproduces the expected results, without any intervening fractions 
in the computation. The details of this computation will be presented
in \cite{qmckay2}.

\section{The quantum McKay correspondence}

It is useful to review some aspects of the McKay correspondence 
that are relevant for our considerations. 
Consider the orbifold $\BC^n/\Gamma$ and its resolution $X$. For the most
part, we will be interested in the cases when $\Gamma$ is abelian.
It is a correspondence between fractional zero-branes
on the orbifold $\BC^n/\Gamma$, $S_a$ ($a$ runs over the irreps of $\Gamma$)
and tautological bundles associated with $2n$-branes, $R^a$. The $R^a$  
extend over all of $\BC^n/\Gamma$ though we usually restrict them to the
singularity (or the exceptional divisor(s)  when the singularity is resolved).
The $S_a$ furnish a basis for $K_c(X)$, the K-theory classes with compact
support on the exceptional divisor(s).

\subsection{Review of $\BC^2/\BZ_{n(k)}$}
This subsection is based on \cite{MM}.
Consider the following orbifold action on $\BC^2$ (with coordinates
$(\phi_1,\phi_2)$):
\begin{equation}
(\phi_1,\phi_2) \rightarrow (\omega \phi_1, \omega^k \phi_2)
\end{equation}
where $\omega=\exp(2\pi i/n)$. The case when $k=(n-1)$ is a supersymmetric
orbifold and the orbifold is uniquely resolved by blowing up $(n-1)$ $\BP^1$'s 
whose intersection matrix is $-1$ times  the $A_{n-1}$ Cartan matrix. 
For general non-supersymmetric $\BC^2/\BZ_{n(k)}$, there is a {\em
minimal resolution} known as the Hirzebruch-Jung resolution. The
resolution consists of $r$ $\BP^1$'s, where $r$ is the number of terms
in the continued fraction expansion of $n/k$:
\begin{equation}
\frac{n}{k} = a_1 - \frac1{a_2-\frac1{a_3 -\frac1{\cdots 1/a_r}}}
\equiv [a_1,a_2,\ldots,a_r]
\end{equation}
where $a_\alpha\geq 2$.  There are other resolutions with more $\BP^1$'s
for which some of the $a_\alpha=1$. The supersymmetric case occurs when
there all the $a_\alpha=2$. One can check that $n/(n-1)=[2^{n-1}]$, 
$3/1 =[3]$ and $5/3=[2,3]$ are minimal. 
The intersection matrix of the $\BP^1$'s is
given by the generalised Cartan matrix 
\begin{equation}
C = \begin{pmatrix} a_1 & -1 & 0 & \cdots & 0 \\
                    -1  & a_2 & -1 & \cdots & 0 \\
                     0  & -1 & a_3 & \ddots & 0 \\
                    \vdots & \vdots & \ddots & \ddots &\vdots \\
                     0  & \cdots & \cdots & & a_r 
\end{pmatrix}
\end{equation}    

\subsection{Fractional zero-branes on $\BC^2/\BZ_{n(k)}$}
 
One can construct boundary states for zero-branes
on the $\BC^2/\BZ_{n(k)}$ orbifold.
Standard methods (analogous to our earlier discussion)
lead to $n$ boundary states which we will call fractional zero-branes
and label them $S^{\textrm{ns}}_I$, $I=1,2,\ldots,n$ where
the superscript stands for non-supersymmetric even though
there are situations where we have a supersymmetric orbifold.
The zero-brane which can move
off the orbifold singularity is given by $\sum_I S^{\textrm{ns}}_I$,

These provide a basis for equivariant K-theory of the orbifold:
\begin{equation}
K_{\BZ_n}(\BC^2) = \BZ \oplus \BZ^{n-1}
\end{equation}
where $\BZ$ denotes the non-fractional zero-brane that can move off the
singularity and $\BZ^{n-1}$, the $(n-1)$ fractional branes that cannot
move off the singularity.

\subsection{The supersymmetric case: the McKay correspondence}
\begin{figure}
\begin{center}
\includegraphics[width=4in]{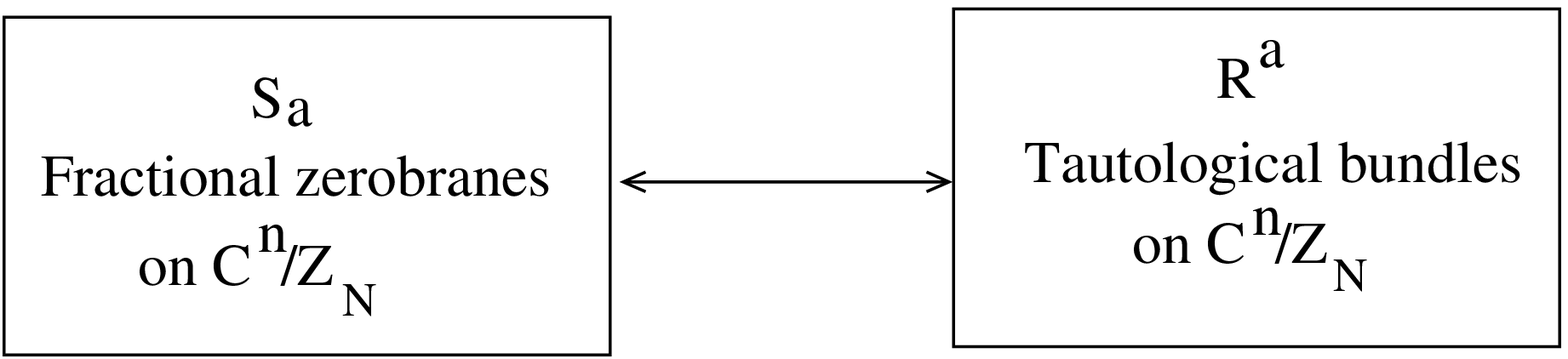}
\caption{The McKay correspondence for $\BC^n/\BZ_N$}
\end{center}
\end{figure}

The McKay correspondence arises when one considers a 
resolution $X$ of the orbifold singularity $\BC^2/\BZ_{n(n-1)}$  
-- for the supersymmetric case, 
we consider the unique crepant (Calabi-Yau) resolution and for the
non-supersymmetric case, we mean the Hirzebruch-Jung resolution. 
One would like to know the precise objects, i.e., coherent sheaves that
correspond to the continuation to large volume of the fractional zero-branes
that we obtain at the orbifold point. We will focus on the
cases where there is a description of the resolution via the GLSM
or equivalently, that the resolved space admits a toric description.

The GLSM for the resolved orbifold will be given by considering $(2+r)$
chiral superfields and $r$ abelian vector multiplets\cite{MM},
where $r$ is the number of terms in the continued fraction
representation of the Hirzebruch-Jung resolution. The orbifold limit is
a special point in the K\"ahler moduli space. Another point of interest
is the large-volume point, which corresponds to the point in the moduli
space where all the $\BP^1$'s that appear in the resolution have been
blown-up to large sizes. Let $D_\alpha$ ($\alpha=1,\ldots,r$)
represent the divisors associated with the $r$ $\BP^1$'s. 

In the supersymmetric case which happens when $r=(n-1)$,
$R^I_{\textrm{ns}}$ turn out to be simple. $(n-1)$ of them are 
given by the line-bundles ${\cal O}(D_\alpha)$ and the last one is
the trivial line-bundle ${\cal O}_X$. These line bundles are
called the {\em tautological bundles} and provide
a basis for $K(X)$, the Grothendieck group of
coherent sheaves on $X$ (which is a non-compact CY two-fold).

The fractional zero-branes furnish a basis for the equivariant K-group
for the orbifold, i.e., $K_{\BZ_n}(\BC^2)$.
In a similar fashion, 
it turns out that the the large-volume analogs of
the fractional zero-branes $S_I^{\textrm{ns}}$ provide a basis for
$K^c(X)$, the K-theory group with compact support. 
One expects the  isomorphism 
$$K_{\BZ_n}(\BC^2)\sim K^c(X)\ .$$
Further, there exists an isomorphism between $K^c(X)$ and $K(X)$.

\subsection{The non-supersymmetric case: the quantum McKay Correspondence}

Martinec and Moore\cite{MM} considered the case of non-supersymmetric orbifolds
and attempted to find the large-volume analogs of the $R^I_{\textrm{ns}}$
here. The {\em natural} candidates are  the line-bundles 
${\cal O}(D_\alpha)$ and the last one is
the trivial line-bundle ${\cal O}_X$. There are $(r+1)$ of them as
in the supersymmetric case with the only problem being that $r+1<n$. So there
are not enough line-bundles to complete the $n$ $R^I_{\textrm{ns}}$
at large-volume. The $r+1$ line-bundles are in one-to-one correspondence
with the so-called {\em
special} representations of $\BZ_n$ in the mathematics
literature\cite{specialreps}.

We now review the resolution of this puzzle as given in \cite{MM}. We
will propose another means of resolving this puzzle in the next
subsection. The framework used in the GLSM that we discussed earlier
where the Hirzebruch-Jung resolution appears in the Higgs branch of the
GLSM.  It is important to note that the Hirzebruch-Jung resolution
is not a crepant one, since $c_1(X)< 0$. 
In the quantum GLSM, the world-sheet FI parameters flow under the worldsheet
renormalisation group\cite{wittenphases}.
The singularities are resolved  in the IR.

The resolution proposed in \cite{MM} is that one {\em must}
include branes from all quantum vacua. In the IR, the theory has two
branches -- the Higgs and the Coulomb branches. The missing branes
were identified with branes that appeared in the Coulomb branch and were
dubbed the {\em Coulomb branch branes}. In analogous fashion, the tautological
bundles in the Hirzebruch-Jung resolution were called the {Higgs branch
branes}. Further aspects were discussed in a subsequent paper\cite{MP}
(see also \cite{Sarkar,narayan}).

\subsection{A different interpretation}
\begin{figure}
\begin{center}
\includegraphics[width=4in]{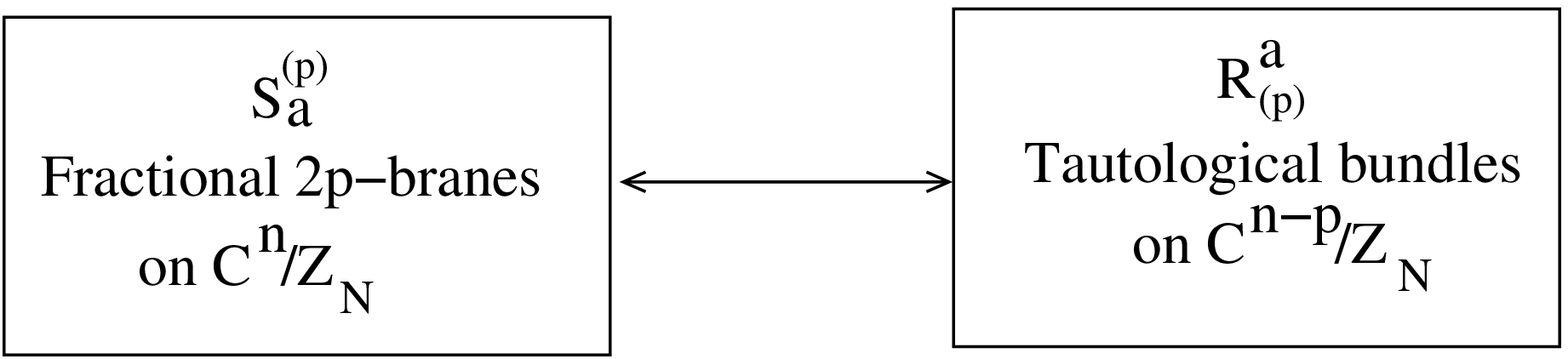}
\caption{The proposed quantum McKay correspondence for $\BC^n/\BZ_N$}
\end{center}
\end{figure}

We now consider a different resolution to the puzzle discussed in the
previous subsection. Our idea is to embed the non-supersymmetric
$\BC^2/\BZ_{n(k)}$ orbifold into a supersymmetric orbifold in one higher
dimension, i.e., $\BC^3/\BZ_{n(k)}$, where we have added a third
coordinate, lets call it $\phi_3$ with the following $\BZ_n$ action:
\begin{equation}
\phi_3 \rightarrow \omega^{l}\ \phi_3,\ \textrm{ where }(l+k+1)=0\textrm{ mod } 2n\ .
\end{equation}
Next, we consider following
fractional two-branes on $\BC^3/\BZ_{n}$:
Impose Neumann boundary conditions on $\phi_3$ and Dirichlet
boundary conditions $\phi_1=\phi_2=0$ on $\phi_1$ and $\phi_2$. 
There will be $n$ such {\em fractional two-branes}
and we will label them 
$S^{(2)}_I$.

Let $\widehat{X}$ be the crepant resolution of $\BC^3/\BZ_{n(k)}$.
It is clear that the the projection $\pi:\widehat{X}\rightarrow X$ is 
obtained by setting $\phi_3=0$.
As discussed in the previous section, there is a problem similar to the
one seen with the non-supersymmetric -- the labels
$I$ corresponding to the {\em special representations} can be obtained
using generalisations of the Euler sequences for the fractional
zero-branes. In fact, one obtains the following when one restricts the
$S^{(2)}_I$ to $X$:
\begin{equation}
S^{(2)}_I\big|_{X}=\left\{ \begin{array}{ll}
                            S^{\textrm{ns}}_I & \textrm{when $I$
corresponds to special representations} \\
0 & \textrm{otherwise}
\end{array}\right. \ .
\end{equation}
This is consistent with our observation
in the $\BC^3/\BZ_3$ example
where  branes which disappeared on restriction are those with support on
the complex line given by $\phi_3\neq0$ and
$\phi_1=\phi_2=0$.\footnote{The resolution of the more general supersymmetric 
$\BC^3/\BZ_{n(k)}$ orbifold requires one to add $(r-1)$ extra fields
and abelian vector multiplets. 
The details of this and related issues will be discussed for more
general cases in  \cite{qmckay2,geometry}. 
In this paper,
we will only provide details for the $\BC^2/\BZ_{3(1)}$ non-supersymmetric
orbifold.}  Thus,
the field $\phi_3$ behaves like an order parameter with $\phi_3=0$
corresponding to the Higgs branch branes and $\phi_3\neq0$ giving rise
to the Coulomb branch branes of \cite{MM}.

\subsubsection{An example -- $\BC^2/\BZ_{3(1)}$}

We have already worked out the large volume continuation of the
fractional two-branes on $\widehat{X}$.
In appendix  \ref{c3intersections},
we have provided the Chern classes for these 
objects. We identified $S^{(2)}_3$ as the Coulomb branch brane. 
What are the
candidates for the $R_{(2)}^I$'s? In $\widehat{X}$, the $\BP^1\in\BP^2$
given by $\phi_3=0$ is to be identified with the $\BP^1$ that appears in
the Hirzebruch-Jung resolution $X$ of $\BC^2/\BZ_{3(1)}$. 
The $R_{\textrm{ns}}^I$ corresponding to special representations
are ${\cal O}_{X}(-1)$  and ${\cal O}_{X}$. 
They are ``dual'' to $S^{(2)}_1\big|_X$
and $S^{(2)}_2\big|_X$.  The natural objects on $\widehat{X}$ are 
the push-forward of the $R$'s on $\BC^2/\BZ_{3(1)}$:
\begin{equation}
R_{(2)}^1\sim 
j_*\big[{\cal O}_{\widehat{X}}(-1) \big]\textrm{ and } 
R_{(2)}^2\sim 
j_*\big[{\cal O}_{\widehat{X}}\big]
\end{equation}
where $j$ is the inclusion map from $X$ to $\widehat{X}$.
The last object, $R^{(2)}_3$ is a little bit more trickier to explain.
We do not present the details here -- it will be presented in
\cite{qmckay2}. Its Chern character as well as the those of the above
$R_{(2)}^I$ can be worked out in the total space using the duality with
the $S^{(2)}_I$. An
important point to emphasise here is that due to the non-compactness of
these D-branes, it is better to work in the total space, which
is the total space of the line bundle 
${\cal O}_{\BP^2}(-3)$ in our case. In fact, this happens to be true
even for fractional zero-branes in cases with several divisors as has
been considered in \cite{skarke,koushik}.

\section{Fractional two-branes and permutation branes}

It is of interest to construct the boundary states in the Gepner model
associated with linear boundary conditions that we considered
in the LG orbifolds. This is what we shall pursue in this section.

\subsection{A conjecture}

Consider the holomorphic
involution  $\pi$ that permutes two fields
\footnote{This is not a symmetry of the Gepner model or the
NLSM but can be made into one by combining it with worldsheet parity.
Thus, this particular involution has been considered in the context of
type IIB orientifolds. }:
$$
\pi:\quad \phi_1 \leftrightarrow  \phi_2
$$
The fixed point(s) of the action is $\phi_1+\phi_2=0,\phi_3=\phi_4=\phi_5=0$
(this is a two-brane which we called $P$ which restricts to a point on
the quintic $Q$)
as well as $\phi_1-\phi_2=0$(this is an eight-brane which restricts to a
four-brane on the quintic). Thus, we see the appearance of the boundary
conditions of Eq. (\ref{quinticbc}) as one of the  fixed point sets
of the holomorphic involution. This suggests that the permutation
branes of ref. \cite{permutation}, 
in particular, those corresponding to $\pi=(12)(3)(4)(5)$ may
be the correct candidate for the boundary states in the Gepner model
that correspond to the boundary conditions given in Eq. (\ref{conda}).
This leads to the following conjecture:

\noindent {\bf Conjecture:} The B-type permutation branes labelled\footnote{We 
choose $S=0$ and $M$, $\widetilde{M}$ to be even.} 
$$|0,0,0,0,M,\widetilde{M}\rangle_{\pi}\quad,\ \pi=(12)(3)(4)(5)$$ 
of ref.\cite{permutation}
are the CFT boundary states for the boundary conditions given in
Eq. (\ref{quinticbc}) associated with the fractional two-branes.  

\subsection{Checks of the conjecture}
\noindent{\bf A first check:}
Recall, that we had obtained {\em five} fractional two-branes at large
volume in section 3. But, we have $25$ boundary states since both
$M$ and $\widetilde{M}$ each take $5$ values.
How can this make sense? In
this regard it is useful to recall that there is a $(\BZ_5)^5$ symmetry
in the minimal models as well as their corresponding LG models which act
as
$$
g_i:\quad \phi_i \rightarrow \omega\ \phi_i\quad,\ g_i^5=1\ ,
$$
where $\omega$ is a non-trivial fifth-root of unity. Focusing on the
$(\BZ_5)^2$ which act on the fields $\phi_1$ and $\phi_2$, we see that
the boundary condition $(\phi_1+\phi_2)=0$ is invariant only under
the simultaneous action $g_1g_2$ while $g_1$ or $g_2$ or the
combination $g_1g_2^{-1}$ act as {\em boundary condition changing
operators}. Thus, the 25 boundary states in the CFT (of
\cite{permutation}) correspond to the five sets of boundary conditions:
\begin{equation}
\phi_1 + \omega^{a} \phi_2 =0 \quad,\ a=0,1,2,3,4.
\label{quinticnewbc}
\end{equation}
Thus, the $\widetilde{M}$ index can be identified with $2a$.

{\bf A second check:} Ref. \cite{permutation} provides the intersection
matrix between the permutation branes (though the normalisation of
the boundary states was not fixed in that paper). The result which we quote here
(after fixing the normalisation) is
the following: the intersection form is {\em independent} of $\widetilde{M}$
and is given by $g(1-g)^3$. It is easy to see that the fractional
two-branes that we propose at large volume also have the same
intersection matrix (see appendix \ref{quinticintersection}). Since the intersection form does
depends on only the Chern character (equivalently, RR charges) of the
coherent sheaves, it will necessarily be independent of the $\widetilde{M}$
label. For instance, the D0-branes obtained from the boundary
conditions, (\ref{quinticnewbc}),  are located at different points on
the quintic. However, their charges are identical.

{\bf A third check:} A last check is to compute the intersection matrix
between the $(12)(3)(4)(5)$ permutation boundary states and the RS
boundary states and show that it equals $-(1-g)^4$ as obtained in
the appendix \ref{quinticintersection} 
by computing the intersection matrix between the RS
vector bundles and coherent sheaves for the fractional two-branes.
This is a little bit more subtle since it involves characters that
are not present in the bulk/closed string partition function. 
This issue has been discussed in a recent paper, so we do not present
any details but refer the reader to it for a more detailed
discussion\cite{Brunner:2005fv}. As is finally shown in
there, the intersection matrix does take the form $-(1-g)^4$ as
predicted by the conjecture.

\subsection{Other permutation branes}

It is natural to see if other permutation branes can be obtained by
considering other linear boundary conditions involving more fields.
One obvious candidate  for the LG boundary conditions associated with 
the permutation brane $(12)(34)(5)$:
\begin{equation}
\phi_1 + \phi_2 = \phi_3 + \phi_4 = \phi_5=0\ .
\label{other}
\end{equation}
This will be a fractional four-brane and the open-string
Witten index computation 
in section 4 for fractional four-branes can be compared with 
the intersection matrices for the $(12)(34)(5)$ branes. The Coulomb
branes have support on the hypersurface $S$ corresponding to
the conditions given in Eq. (\ref{other}). This intersects the $\BP^4$
on a $\BP^1$. The Coulomb branes will be given by ${\cal X}_S\otimes
{\cal O}(-1)$ and  ${\cal X}_S\otimes {\cal O}$, 
where ${\cal X}_S$ is the sheaf with support
on the hypersurface $S$ (and Chern class $J^2/5$ reflecting the
fractional charge) 
and $({\cal O}, {\cal O}(1))$ are the
tautological bundles on $\BP^1$. Thus, their Chern classes will be
$$ 
\frac{J^2}5- \frac{J^3}5 \ , \textrm{ and } \frac{J^2}5 \ .
$$
This is consistent with the Chern classes of the permutation branes
labelled $V_3$ and $V_4$ in Eq. (6.13) of \cite{Brunner:2005fv}.
These restrict to a two-brane (of minimal charge) on the quintic.
The Higgs branes will be related to Euler sequences on $\BP^2$. The
other branes given in the aforementioned reference also seem to 
fit this. So this also passes a similar set of checks. Thus,
for the quintic, one is able to obtain objects which reproduce the
full spectrum of charges -- the fractional two-branes providing the
zero-brane and the fractional four-brane providing the
two-brane.

Another interesting boundary condition in the LG orbifold is the fractional
four-brane given by 
$$
\phi_1 + \phi_2 + \phi_3 = \phi_4 = \phi_5=0\ .
$$
This involves three fields and seems to be related to the 
permutation brane $(123)(4)(5)$. However, this runs into trouble in
the first check\footnote{There is a subtlety even in the earlier
examples with regard to other permutation branes with $L\neq0$. It has
been shown by the authors of \cite{Brunner:2005fv} that the CFT states
with $L=1$ correspond to (in the LG) to quadratic factors chosen in 
a particular order. However, in the LG there doesn't seem to be any
reason to prefer the ordering. These issues have been discussed in
\cite{Brunner:2005fv}. We thank Ilka Brunner for drawing our attention
to this interesting issue.}
correspond to . The permutation branes constructed in
\cite{permutation} do not have enough labels to account for the $125$
boundary states that one anticipates by considering the action of the
symmetries on this boundary condition. In this regard, we wish to point
out that for cyclic orbifolds\cite{Klemm,BHS}, 
$\BZ_\lambda$ with $\lambda >2$, there
is a fixed point resolution problem, related to the fact that the
different primaries in the orbifold theory have the same character
(see section 5 of \cite{BHS}).
This also suggests that the set of permutation branes
given in \cite{permutation} may not be minimal and their resolution will
provide us with additional boundary states that may account for the
$125$ boundary states that we predict. We hope to discuss this issue
in a future publication.

\section{Conclusion and Summary}

We first summarise the main results of this paper. 
\begin{itemize}
\item We have provided evidence that the new fractional branes
obtained in \cite{Ashok} using the method of matrix factorisation
and boundary fermions are the same as those given by the boundary conditions
proposed in \cite{gjs}. The methods used in this paper also lead
to sequences for the new fractional branes which carry more
information than Chern classes/RR charges. As a consequence, we
show that fractional $2p$-branes constructed in the non-compact 
Calabi-Yau restrict to the compact Calabi-Yau hypersurface as well
defined objects.
\item We also provided evidence that a sub-class of the permutation branes
proposed in \cite{permutation} are related to the new fractional
branes. This result has also been independently
obtained by Brunner and Gaberdiel recently\cite{Brunner:2005fv}.
In particular, a detailed discussion of the permutation branes and
the intersection matrices amongst them has been provided in their
paper.
\item By embedding non-supersymmetric orbifolds into supersymmetric
orbifolds in one higher dimension, we propose a quantum McKay
correspondence that relates fractional $2p$-branes on $\BC^n/\BZ_N$
to  (the push-forward of) tautological bundles on
$\BC^{n-p}/\BZ_N$ branes in supersymmetric orbifolds. We also
provide an
alternate explanation  to the ``missing'' branes discussed in \cite{MM}.
\end{itemize}

This paper has dealt with, in many ways, the easiest of the examples.
Possible generalisations include the consideration of other examples 
involving
weighted projective spaces -- these typically involve many K\"ahler
moduli and the sequences that we proposed in this paper for the
new fractional branes will have a more complicated realisation. We
plan to discuss this in \cite{qmckay2}. 

Another class of boundary
conditions that can appear in these examples are those that are not
linear. For instance, consider the boundary condition $(\phi_1 +
\phi_2^2)=0$ that is consistent with the $G=0$ condition. How does
one construct the boundary states corresponding to such a boundary
condition? An added complication is that
they do not seem to be minimal submanifolds.
What are their intersection forms? In fact, it has been
argued in \cite{Brunner:2005fv} that in more complicated examples,
the addition of the analogue of the fractional two-branes and
four-branes that we have considered do not give rise to all possible
RR charge vectors. Thus, one may be forced to deal with such
boundary conditions. However the Gepner model 
for such examples typically involve minimal models with
even level $k$. Here, even the RS boundary states are not minimal in the
Cardy sense. So the issue is somewhat clouded by the need to {\em
resolve} the boundary states.

Coming back to the case of the quintic -- while it is indeed
satisfying to find objects in the Gepner model/LG orbifold which
provide all RR charges that appear on the quintic, it cannot
be the end of the story. The RS boundary states were related to
spherical objects (spherical in the sense that these objects have only
${\rm Hom}(E,E) \equiv {\rm Ext}^3(E,E) \neq 0$)
on the quintic even though they did not span
the lattice of RR charges. One would like to know, if there exists
a basis of, say, four spherical objects on the quintic that give rise to all 
possible charges via bound state formation. In the framework of boundary
fermions and matrix factorisation 
the superpotentials on these branes have been computed 
and their relationship to obstruction theory has been discussed. 
It would be interesting to see if
those computations agree with an extension of the superpotential
computation carried out for the RS branes in the LG model
in\cite{DGJT} to the case of the fractional two-branes and four-branes.

In this paper, we have shown a parallel between fractional $2p$-branes
on supersymmetric orbifolds and fractional zero-branes on
non-supersymmetric orbifolds of dimensional lower by one. 
The Coulomb branes in the
non-supersymmetric orbifolds have been identified with the minima of the
quantum superpotential of twisted chiral superfields.  The Coulomb
branes, as we refer to them, among the fractional two-branes 
in our construction are associated with a chiral
superfield. It appears possible to make the connection between 
the two situations
precise by applying the Hori-Vafa map\cite{HV}, 
which relates chiral superfields
to twisted chiral superfields. 
This issue as well as an explanation for
the change of basis proposed in \cite{MP} will be discussed in
\cite{geometry}.

Our reinterpretation of the quantum McKay correspondence will be
useful for non-supersymmetric orbifolds in higher dimensions
such as $\BC^3/\BZ_n$ \cite{Sarkar,narayan} where there is no 
analogue of the
Hirzebruch-Jung resolution via partial fractions. There is also
the problem of terminal singularities that can appear. 
At least in cases where the higher dimensional supersymmetric orbifold
admits a crepant resolution, one will be able to use the
embedding to study aspects of the non-supersymmetric orbifold
such as the possible end-points of tachyon condensation.\\

\noindent {\bf Acknowledgments:} We thank A. Adams, Shiraz Minwalla,
Kapil Paranjape,
K. Ray, T. Sarkar and T. Takayanagi for useful conversations. S.G.
would like to thank Shiraz Minwalla and the Physics Department of
Harvard University for hospitality where some of this work was done
during summer 2004. T.J would like to thank the Dept. of Theoretical
Physics, TIFR for hospitality where part of this work was done.
B.E. would like to thank Dept. of Theoretical Physics, TIFR for hospitality
during the summer of 2004. T.J. would like to thank N. Nitsure and V. Srinivas
for several explanations and 
useful discussions that were important to understanding 
the mathematics relevant to this paper.

\appendix

\section{Intersection matrices for the quintic}
\label{quinticintersection}

The Chern character of the {\em RS branes} on the Fermat quintic $Q\in \BP^4$
are obtained from the generalised Euler sequences given in Eq.
(\ref{generalisedeuler}). One obtains
\begin{eqnarray*}
\ch({\cal M}_0) &=& 1 \\
\ch({\cal M}_1) &=& -4 + J + \frac{J^2}{2} + \frac{J^3}{6} \\
\ch({\cal M}_2) &=& 6 - 3\,J - \frac{J^2}{2} + \frac{J^3}{2} \\
\ch({\cal M}_3) &=& -4 + 3\,J - \frac{J^2}{2} - \frac{J^3}{2} \\
\ch({\cal M}_4) &=& 1 - J + \frac{J^2}{2} - \frac{J^3}{6} 
\end{eqnarray*}
In the above $J$ generates $H^2(Q,\BR)$ with the normalisation chosen to be
$\langle J^3 \rangle_Q = 5 \langle J^4 \rangle_{\BP^4}=5$.
The Chern character for the {\em new fractional branes} are given
in terms of the sequences that we proposed in Eqs. (\ref{neweuler}). The only
additional input is our proposal that $\ch({\cal X}_P)=\frac{J^3}{5}$.
\begin{eqnarray*}
\ch({\cal F}_0) &=& 1 - \frac{J^3}{5} \\
\ch({\cal F}_1) &=&  -3 + J + \frac{J^2}{2} - \frac{J^3}{30}\\
\ch({\cal F}_2) &=& 3 - 2\,J + \frac{7\,J^3}{15} \\
\ch({\cal F}_3) &=& -1 + J - \frac{J^2}{2} - \frac{J^3}{30} \\
\ch({\cal F}_4) &=&  \frac{-J^3}{5}
\end{eqnarray*} 

The Euler form on the quintic is defined as
$$
\chi(E,F) = \int_Q \ch(E)^*\ \ch(F)\ \textrm{Td}(Q)\ ,
$$
A standard computation (easily done using symbolic manipulation programs)
using the above formula leads to 
the following intersection matrices are:
\begin{equation}
\chi({\cal M}_\mu, {\cal M}_\nu) = \begin{pmatrix}
0& 5& -10& 10& -5 \\ 
-5& 0& 5& -10& 10\\ 
10& -5& 0& 5& -10 \\
-10& 10& -5& 0& 5\\
5& -10& 10& -5& 0
\end{pmatrix}\longleftrightarrow -(1-g)^5
\end{equation}
\begin{equation}
\chi({\cal M}_\mu, {\cal F}_\nu) = \begin{pmatrix}
-1& 4& -6& 4& -1\\
-1& -1& 4& -6& 4\\
4& -1& -1& 4& -6\\
-6& 4& -1& -1& 4\\
4& -6& 4& -1& -1
\end{pmatrix} \longleftrightarrow -(1-g)^4
\end{equation}
\begin{equation}
\chi({\cal F}_\mu, {\cal F}_\nu) = \begin{pmatrix}
0& 1& -3& 3& -1\\
-1& 0& 1& -3& 3\\
3& -1& 0& 1& -3\\
-3& 3& -1& 0& 1\\
1& -3& 3& -1& 0
\end{pmatrix} \longleftrightarrow g(1-g)^3
\end{equation}
where we have rewritten the matrices in terms of the generator $g$ of the 
quantum $\BZ_5$ symmetry.  The last formula coincides
with the intersection matrix
computed between the $L=0$ permutation branes for $\pi=(12)345$ given
in the appendix of \cite{permutation}.

\section{Intersection matrices for $\BC^3/\BZ_3$}
\label{c3intersections}

In this appendix, we present the computation of various intersection
matrices for the coherent sheaves given by the sequences written out
for fractional zero and two-branes. We present the Chern classes after
{\em restriction} to the $\BP^2$.
The Chern classes for the fractional zero-branes are 
given from the Euler sequences for $\BP^2$
\begin{eqnarray}
S^{(0)}_1 &=& 1 \\
S^{(0)}_2 &=& -2 + J + \frac{J^2}{2} \\
S^{(0)}_3 &=& 1 - J + \frac{J^2}{2}
\end{eqnarray}
The Chern classes for the fractional two-branes are
\begin{eqnarray}
S^{(2)}_1 &=& 1 - \frac{J}{3} - \frac{(2m+3)J^2}{6} \\
S^{(2)}_2 &=& -1 + \frac{2\,J}{3}-\frac{(m+1)J^2}{3} \\
S^{(2)}_3 &=& -\frac{J}{3} - \frac{(2m+1)J^2}{6}
\end{eqnarray}

The various Euler forms given below are obtained using the formula:
$$
\chi(E,F) = \int_{\BP^2} \ch(E)^*\ \ch(F)\ \textrm{Td}(\BP^2)\  
$$
\begin{equation}
\chi(S^{(0)},S^{(0)}) = \begin{pmatrix}
                        1& 0& 0\\ -3& 1& 0 \\ 3& -3& 1 
                        \end{pmatrix}
\end{equation}

\begin{equation}
\chi(S^{(2)},S^{(0)}) = \frac13\begin{pmatrix}
                        -m+3& 2m+1 & -m-1 \\
                        -m-7&2m+6&-m-2 \\
                        -m+1& 2m-1& -m
                        \end{pmatrix}
\end{equation}
\begin{equation}
\chi(S^{(0)},S^{(2)}) = \frac13\begin{pmatrix}
                        -m&-1-m&-m-2 \\
                        2m-2 & 2m+3& 2m+5 \\
                        -m+5& -m-5 & -m-3  
                        \end{pmatrix}
\end{equation}
\begin{equation}
\chi(S^{(2)},S^{(2)}) = \frac19 \begin{pmatrix}
                         -6m-1 & -1 & -3m -7 \\
                          -10 & 6m+11 & 3m+8 \\
                          -3m +2 & 3m-1 & -1
                        \end{pmatrix}
\end{equation}
For bundles on a Calabi-Yau manifold these are antisymmetric. However, for
bundles on divisors or more generally sheaves, the intersection form is
obtained by explicitly antisymmetrising the above(as explained in
\cite{dfr1}), i.e., let
$$
I(E',E)  \equiv \chi(E',E) - \chi(E,E')
$$
for any two sheaves $E$ and $E'$. In our case, we then get
\begin{eqnarray}
{\cal I}^{0,0}\equiv I(S^{(0)},S^{(0)}) &=& -(1-g)^3 \nonumber \\
{\cal I}^{0,2}\equiv I(S^{(0)},S^{(2)}) &=& -(1-g)^2 \\
{\cal I}^{2,2}\equiv I(S^{(2)},S^{(2)}) &=& g(1-g)\nonumber  
\end{eqnarray}
Note that the dependence on $m$ disappears in the intersection matrix
and thus we cannot fix its value. However, this is to be expected since
a change in $m$ is obtained by twisting by ${\cal O}(1)$, which is the
monodromy at large volume. Since these sheaves are obtained via analytic
continuation, the ambiguity in $m$ is related to the possibility of
choosing paths from the orbifold point to large volume which differ by
a path around the large-volume point.
The fact that the above matrix has {\em integer} entries
implies that the DSZ charge quantisation condition is satisfied.
Thus it is correct to assume that the fractional two-branes do carry
fractional two-brane RR charge and not integral as suggested in
\cite{tadashi}
(in particular see appendix A, Eq. (A31)) suggests.

\section{Cohomology and K-theory computations}

\subsection{Relevant cohomology groups for ${\cal O}_{\BP^2}(-3)$}
\label{cohomology}

We consider a spacetime of the form $\BR\times M$, where $\BR$
represents the time direction and $M$ is non-compact. 
Let $N$ be the boundary of $M$. The D-brane charges take values in
the relative cohomology $H^{*}(M,N;\BZ)$\cite{moorewitten}. 
These can be computed by
considering the long-exact sequence in cohomology:
\begin{equation}
\cdots \rightarrow
H^p(M;\BZ) \stackrel{j}{\rightarrow}
H^p(N;\BZ) \rightarrow
H^{p+1}(M,N;\BZ) \stackrel{i}{\rightarrow}
H^{p+1}(M;\BZ) \rightarrow \cdots
\end{equation}
The map $j$ corresponds to restricting $p$-forms on $M$ to the boundary
$N$.
Let us choose $M$ to be $\BC^3/\BZ_3$ (or equivalently ${\cal
O}_{\BP^2}(-3)$). Then, one has $N=\partial M= S^5/\BZ_3$. 
One has the following non-vanishing cohomologies for $N$ and $M$.
\begin{eqnarray}
H^4(S^5/\BZ_3;\BZ) &=& H^2(S^5/\BZ_3;\BZ) = \BZ_3\quad, 
\nonumber \\
H^0(S^5/\BZ_3;\BZ) &=& H^5(S^5/\BZ_3;\BZ) = \BZ\quad,\quad
\\
H^{2p}({\cal O}_{\BP^2}(-3);\BZ) &=& H^{2p}(\BP^2;\BZ)=\BZ \quad,\quad
\textrm{for } p=0,1,2\quad, \nonumber
\end{eqnarray}
Using the above data, we obtain the long-exact sequence breaks into
four shorter sequences:
\begin{eqnarray}
0\rightarrow H^{0}({\cal O}_{\BP^2}(-3), S^5/\BZ_3;\BZ) \rightarrow \BZ
\stackrel{j}{\rightarrow} \BZ  \rightarrow H^{1}({\cal
O}_{\BP^2}(-3), S^5/\BZ_3;\BZ) \rightarrow 0 \\
0\rightarrow H^{2}({\cal O}_{\BP^2}(-3), S^5/\BZ_3;\BZ) \rightarrow \BZ
\stackrel{j}{\rightarrow} \BZ_3  \rightarrow H^{3}({\cal
O}_{\BP^2}(-3), S^5/\BZ_3;\BZ) \rightarrow 0 \\
0\rightarrow H^{4}({\cal O}_{\BP^2}(-3), S^5/\BZ_3;\BZ) \rightarrow \BZ
\stackrel{j}{\rightarrow} \BZ_3  \rightarrow H^{5}({\cal
O}_{\BP^2}(-3), S^5/\BZ_3;\BZ) \rightarrow 0 \\
0\rightarrow \BZ \rightarrow H^{6}({\cal O}_{\BP^2}(-3), S^5/\BZ_3;\BZ) 
\rightarrow 0
\end{eqnarray}
In the first equation above, the map $j$ corresponds to restricting
(constant) functions on  ${\cal O}_{\BP^2}(-3)$ to the boundary
$S^5/\BZ_3$. Clearly, this is an isomorphism and hence we obtain
$H^{0}({\cal O}_{\BP^2}(-3), S^5/\BZ_3;\BZ)$ and   $H^{1}({\cal
O}_{\BP^2}(-3), S^5/\BZ_3;\BZ)$ both vanish. 
The last equation implies that $H^{6}({\cal O}_{\BP^2}(-3),
S^5/\BZ_3;\BZ)=\BZ$. It can be shown that all odd (relative) cohomologies
vanish.  Thus, the second and third equations above reduce to
$$
0\rightarrow H^{2p}({\cal O}_{\BP^2}(-3), S^5/\BZ_3;\BZ) 
\stackrel{a}{\rightarrow} \BZ
\stackrel{j}{\rightarrow} \BZ_3  \rightarrow 0\quad (\textrm{for }p=1,2)
$$
This implies that both above relative cohomologies are $\BZ$ with the
first map $a$ representing multiplication by $3$. Thus, one has 
the following result:
\begin{equation}
\boxed{
H^{2p}({\cal O}_{\BP^2}(-3), S^5/\BZ_3;\BZ) = \BZ \quad (\textrm{for
}p=1,2,3)
}
\end{equation}

\subsection{The K-theory computation }
\label{ktheory}
The K-theory computation in the relative case of a manifold with
boundary that parallels the computation of \ref{cohomology} may
also be carried out. In the case at had we are interested in computing
the relative K-groups of the total space of the bundle
${\cal O}(-3)$ on $\BP^2$ with respect to the boundary $S^5/BZ_3$.
The computation is done by examining, as is typical in such situations, a  
six-term exact sequence (a consequence of collapsing a long exact sequence of 
K-groups using Bott periodicity) and using the known results on the K-group of 
$\BP^2$ to compute the relative K-groups of interest. A useful reference for
the computations of this section is \cite{karoubi}.
 
We will take as the manifold $M$, the disc bundle $D(E)$ related to the bundle 
$E$ defined by taking 
all vectors $v$ in $E$ such that their inner product 
(defined with respect to some appropriate
Riemannian metric on $E$) $\langle v,v\rangle \leq 1$. 
The boundary of $D(E)$ will be sphere 
bundle $S(E)$ made of vectors $v$ in $E$ such that $\langle v,v\rangle = 1$. 
By a standard argument, the relative K-groups
$K(D(E), S(E))$ may be identified with the K-groups $K(E, E_0)$ where $E_0$
is the complement of the zero-section of $E$.

To compute the K-groups $K(M, \partial M)$ for a compact manifold $M$ 
with boundary $\partial M$ we
may use the following six-term exact sequence:
\begin{eqnarray}
K^0(M) \rightarrow K^0(\partial M) \rightarrow K^1(M, \partial M) 
\rightarrow K^1(M) \rightarrow \\ \nonumber
\rightarrow K^1(\partial M) \rightarrow K^0(M, \partial M) \rightarrow K^0(M).
\end{eqnarray}
In our case $M=D(E)$, $\partial M = S(E)$. For the case of the disc bundle 
since it is a deformation retract of 
$E$ itself  and the K-group of a bundle is isomorphic to the K-group of 
the base, we have 
$K(D(E))=K(\BP^2)$. To compute $K(S)$ we may compute it knowing the cohomology 
of $S(E)=S^5/\BZ_3$. Though strictly speaking we should use the machinery of
the Atiyah-Hirzebruch spectral sequence, it appears here to provide no 
surprises. We therefore get a $K^0(S(E))$ and $K^1(S(E))$ that is isomorphic
to the cohomology.
 
We may now do the computation, using the data on the K-groups of 
$S(E)$ and $K(\BP^2)$ (in particular, $K^1(\BP^2)=0$, since all the odd
cohomologies of $\BP^2$ are zero)
to obtain the following shorter exact sequence:
\begin{eqnarray}
0 \rightarrow K^1(S^5/\BZ_3) \rightarrow K^0({\cal O}(-3),S^5/\BZ_3) 
\rightarrow K^0(\BP^2) \rightarrow K^0(S^5/\BZ_3) \rightarrow 0.
\end{eqnarray}
Using the data
\begin{eqnarray}
K^0(\BP^2) = \BZ \oplus \BZ \oplus \BZ;
K^0(S^5/\BZ_3) = \BZ \oplus \BZ_3 \oplus \BZ_3; K^1(S^5/\BZ_3) = \BZ
\end{eqnarray}
we obtain the result 
\begin{equation}
K^0(M, \partial M) = \BZ \oplus \BZ \oplus \BZ
\end{equation}
Note that one $Z$ is in the kernel of the map from $K^0(M, \partial M)$
to $K^0(P^2)$. Two of the $Z$ factors in $K^0(M, \partial M)$ 
form part of the sequence of the type 
\begin{eqnarray}
0 \rightarrow  \BZ \stackrel {n} \rightarrow \BZ \rightarrow \BZ_n 
\rightarrow 0.
\end{eqnarray}
where $n=3$ in our case.
What this shows is that if $[J']$ is the generator of $K^0(P^2)$
and $[J]$ is the generator of $K^0(M, \partial M)$ then 
$[J]\sim 3 [J']$. 
If we take the standard unit of 2-brane charges to be given by
$[J]$, then a $1/3$ charge in the $[J]$ basis becomes an integer charge
in the $[J']$ basis.


\newpage
\begin{thebibliography}{99}
\bibitem{quintic} 
I.~Brunner, M.~R.~Douglas, A.~E.~Lawrence and C.~Romelsberger,
``D-branes on the quintic,''
JHEP {\bf 0008} (2000) 015
[arXiv:hep-th/9906200].
\bibitem{gjs}
S.~Govindarajan, T.~Jayaraman and T.~Sarkar,
``Worldsheet approaches to D-branes on supersymmetric cycles,''
Nucl.\ Phys.\ B {\bf 580}, 519 (2000)
[arXiv:hep-th/9907131].
\bibitem{mikeictp} M.~R.~Douglas, 
``Lectures on D-branes on Calabi-Yau manifolds,''
{\it Lectures at the 2001 ICTP Spring School on Superstrings and Related
Matters}, online proceedings are 
available at the URL\\
\verb|http://www.ictp.trieste.it/~pub_off/lectures/vol7.html|
\bibitem{rs}
A.~Recknagel and V.~Schomerus,
``D-branes in Gepner models,''
Nucl.\ Phys.\ B {\bf 531}, 185 (1998)
[arXiv:hep-th/9712186].
\bibitem{dfr1}
M.~R.~Douglas, B.~Fiol and C.~R\"omelsberger,
``Stability and BPS branes,'' {\tt hep-th/0002037}. 
\bibitem{dfr2}
M.~R.~Douglas, B.~Fiol and C.~Romelsberger,
``The spectrum of BPS branes on a noncompact Calabi-Yau,''
arXiv:hep-th/0003263.
\bibitem{dd}
D.~E.~Diaconescu and M.~R.~Douglas,
``D-branes on stringy Calabi-Yau manifolds,''
arXiv:hep-th/0006224.
\bibitem{mckay}
S.~Govindarajan and T.~Jayaraman,
``D-branes, exceptional sheaves and quivers on Calabi-Yau manifolds: From
Mukai to McKay,''
Nucl.\ Phys.\ B {\bf 600}, 457 (2001)
[arXiv:hep-th/0010196].
\bibitem{tomas}
A.~Tomasiello,
``D-branes on Calabi-Yau manifolds and helices,''
JHEP {\bf 0102}, 008 (2001)
[arXiv:hep-th/0010217].
\bibitem{mayr}
P.~Mayr,
``Phases of supersymmetric D-branes on Kaehler manifolds and the McKay
correspondence,''
JHEP {\bf 0101} (2001) 018
[arXiv:hep-th/0010223].
\bibitem{Lerche:2001vj}
W.~Lerche, P.~Mayr and J.~Walcher,
``A new kind of McKay correspondence from non-Abelian gauge theories,''
arXiv:hep-th/0103114.
\bibitem{kapustinli}
  A.~Kapustin and Y.~Li,
  ``D-branes in Landau-Ginzburg models and algebraic geometry,''
  JHEP {\bf 0312}, 005 (2003)
  [arXiv:hep-th/0210296]. \\
  ``Topological correlators in Landau-Ginzburg models with
  boundaries,''
  Adv.\ Theor.\ Math.\ Phys.\  {\bf 7}, 727 (2004)
  [arXiv:hep-th/0305136]. \\
  ``D-branes in topological minimal models: The Landau-Ginzburg
  approach,''
  JHEP {\bf 0407}, 045 (2004)
  [arXiv:hep-th/0306001].
\bibitem{Ashok}
S.~K.~Ashok, E.~Dell'Aquila and D.~E.~Diaconescu,
``Fractional branes in Landau-Ginzburg orbifolds,''
arXiv:hep-th/0401135. 
\bibitem{horiwalcher}
K.~Hori and J.~Walcher,
``F-term equations near Gepner points,''
JHEP {\bf 0501}, 008 (2005)
[arXiv:hep-th/0404196]. \\
``D-branes from matrix factorizations,''
Comptes Rendus Physique {\bf 5}, 1061 (2004)
[arXiv:hep-th/0409204].
\bibitem{lerchewalcher}
 I.~Brunner, M.~Herbst, W.~Lerche and B.~Scheuner,
``Landau-Ginzburg realization of open string TFT,''
arXiv:hep-th/0305133. \\
I.~Brunner, M.~Herbst, W.~Lerche and J.~Walcher,
``Matrix factorizations and mirror symmetry: The cubic curve,''
arXiv:hep-th/0408243.
\bibitem{walcher} 
J.~Walcher,
``Stability of Landau-Ginzburg branes,''
arXiv:hep-th/0412274.
\bibitem{Ashok2}
S.~K.~Ashok, E.~Dell'Aquila, D.~E.~Diaconescu and B.~Florea,
``Obstructed D-branes in Landau-Ginzburg orbifolds,''
arXiv:hep-th/0404167. 
\bibitem{orlov}
D.~Orlov,
``Triangulated categories of singularities and D-branes in
Landau-Ginzburg models,''
arXiv:math.ag/0302304. \\
``Triangulated categories of singularities and equivalences between
Landau-Ginzburg models,''
arXiv:math.ag/0503630. \\
``Derived categories of coherent sheaves and triangulated categories of
singularities,''
arXiv:math.ag/0503632. 
\bibitem{bundlesglsm}
S.~Govindarajan and T.~Jayaraman,
``Boundary fermions, coherent sheaves and D-branes on Calabi-Yau  manifolds,''
Nucl.\ Phys.\ B {\bf 618}, 50 (2001)
[arXiv:hep-th/0104126].
\bibitem{romel1}
  C.~Romelsberger,
``(Fractional) intersection numbers, tadpoles and anomalies,''
arXiv:hep-th/0111086.
\bibitem{BCR}
M.~Billo, B.~Craps and F.~Roose,
``Orbifold boundary states from Cardy's condition,''
JHEP {\bf 0101} (2001) 038
[arXiv:hep-th/0011060].
\bibitem{MM}
E.~J.~Martinec and G.~W.~Moore,
``On decay of K-theory,''
arXiv:hep-th/0212059.
\bibitem{permutation}
A.~Recknagel,
``Permutation branes,''
JHEP {\bf 0304} (2003) 041
[arXiv:hep-th/0208119]. 
\bibitem{Brunner:2005fv}
I.~Brunner and M.~R.~Gaberdiel,
``Matrix factorisations and permutation branes,''
arXiv:hep-th/0503207.
\bibitem{wittenphases}
E.~Witten, ``Phases of $N=2$ Theories in Two Dimensions,''
Nucl. Phys. {\bf B403} (1993) 159 [arXiv:hep-th/9301042].
\bibitem{glsm}
S.~Govindarajan, T.~Jayaraman and T.~Sarkar,
``On D-branes from gauged linear sigma models,''
Nucl.\ Phys.\ B {\bf 593}, 155 (2001)
[arXiv:hep-th/0007075].
\bibitem{gzero}
S.~Govindarajan and T.~Jayaraman,
``On the Landau-Ginzburg description of boundary CFTs and special  Lagrangian
submanifolds,''
JHEP {\bf 0007} (2000) 016
[arXiv:hep-th/0003242]. \\
K.~Hori, A.~Iqbal and C.~Vafa,
``D-branes and mirror symmetry,''
arXiv:hep-th/0005247. 
\bibitem{bd}
I.~Brunner and J.~Distler,
``Torsion D-branes in nongeometrical phases,''
Adv.\ Theor.\ Math.\ Phys.\  {\bf 5} (2002) 265
[arXiv:hep-th/0102018].
\bibitem{diacgom}
D.~E.~Diaconescu and J.~Gomis,
``Fractional branes and boundary states in orbifold theories,''
JHEP {\bf 0010} (2000) 001
[arXiv:hep-th/9906242].
\bibitem{Gaberdiel:1999ch}
M.~R.~Gaberdiel and B.~J.~Stefanski,
``Dirichlet branes on orbifolds,''
Nucl.\ Phys.\ B {\bf 578} (2000) 58
[arXiv:hep-th/9910109].
\bibitem{tadashi}
T.~Takayanagi,
``Holomorphic tachyons and fractional D-branes,''
Nucl.\ Phys.\ B {\bf 603} (2001) 259
[arXiv:hep-th/0103021].
\bibitem{BDFLM}
M. Bertolini, P. Di Vecchia, M. Frau, A. Lerda, R. Marotta and I.
Pesando,
"Fractional D-branes and their gauge duals"
JHEP 0102 (2001) 014
[arXiv:hep-th/0011077] \\
M.~Bertolini, P.~Di Vecchia, M.~Frau, A.~Lerda and R.~Marotta,
``N = 2 gauge theories on systems of fractional D3/D7 branes,''
Nucl.\ Phys.\ B {\bf 621} (2002) 157
[arXiv:hep-th/0107057].
\bibitem{gaberdiel}
M.~R.~Gaberdiel and H.~Klemm,
``N = 2 superconformal boundary states for free bosons and fermions,''
Nucl.\ Phys.\ B {\bf 693} (2004) 281
[arXiv:hep-th/0404062].
\bibitem{specialreps} O. Riemenschneider, ``Special representations and
the two-dimensional McKay correspondence,'' Hokkaido Math.
Journal {\bf XXXII} (2003) 317-333.
\bibitem{MP}
G.~W.~Moore and A.~Parnachev,
``Localized tachyons and the quantum McKay correspondence,''
JHEP {\bf 0411} (2004) 086
[arXiv:hep-th/0403016].
\bibitem{Sarkar}
T.~Sarkar,
``On localized tachyon condensation in $\BC^2/\BZ_n$ and $\BC^3/\BZ_n$,''
Nucl.\ Phys.\ B {\bf 700} (2004) 490
[arXiv:hep-th/0407070].
\bibitem{narayan}
D.~R.~Morrison, K.~Narayan and M.~R.~Plesser,
``Localized tachyons in $\BC^3/\BZ_N$,''
JHEP {\bf 0408}, 047 (2004)
[arXiv:hep-th/0406039].\\
D.~R.~Morrison and K.~Narayan,
``On tachyons, gauged linear sigma models, and flip transitions,''
JHEP {\bf 0502}, 062 (2005)
[arXiv:hep-th/0412337].
\bibitem{qmckay2} Bobby Ezhuthachan, Suresh Govindarajan and T. Jayaraman
(to appear).
\bibitem{geometry} S. Govindarajan, T. Jayaraman  and
Tapobrata Sarkar (to appear).
\bibitem{skarke}
X.~De la Ossa, B.~Florea and H.~Skarke,
``D-branes on noncompact Calabi-Yau manifolds: K-theory and monodromy,''
Nucl.\ Phys.\ B {\bf 644} (2002) 170
[arXiv:hep-th/0104254].
\bibitem{koushik}
S.~Mukhopadhyay and K.~Ray,
``Fractional branes on a non-compact orbifold,''
JHEP {\bf 0107}, 007 (2001)
[arXiv:hep-th/0102146].
\bibitem{Klemm}
A.~Klemm and M.~G.~Schmidt,
``Orbifolds By Cyclic Permutations Of Tensor Product Conformal Field
Theories,''
Phys.\ Lett.\ B {\bf 245} (1990) 53. \\
J.~Fuchs, A.~Klemm and M.~G.~Schmidt,
``Orbifolds by cyclic permutations in Gepner type superstrings and in
the corresponding Calabi-Yau manifolds,''
Annals Phys.\  {\bf 214} (1992) 221.
\bibitem{BHS}
L.~Borisov, M.~B.~Halpern and C.~Schweigert,
``Systematic approach to cyclic orbifolds,''
Int.\ J.\ Mod.\ Phys.\ A {\bf 13} (1998) 125
[arXiv:hep-th/9701061].
\bibitem{DGJT}
M.~R.~Douglas, S.~Govindarajan, T.~Jayaraman and A.~Tomasiello,
``D-branes on Calabi-Yau manifolds and superpotentials,''
Commun.\ Math.\ Phys.\  {\bf 248}, 85 (2004)
[arXiv:hep-th/0203173].
\bibitem{HV}
K..~Hori and C.~Vafa, ``Mirror symmetry,'' arXiv:hep-th/0002222.
\bibitem{moorewitten}
G.~W.~Moore and E.~Witten,
``Self-duality, Ramond-Ramond fields, and K-theory,''
JHEP {\bf 0005} (2000) 032
[arXiv:hep-th/9912279].
\bibitem{karoubi} M. Karoubi, {\em K-Theory: An Introduction},
Springer-Verlag, 1978.
\end{thebibliography}
\end{document}